# The Roles of Familiarity Design in Active Ageing

# [A Book Draft]

By:

Zhengxiang Pan, Jedi

Version: 22-Jan-2016

# Table of Contents

### Keywords

Familiarity; Gerontology; Gerontechnology; Elderly; Reduce their distress; Routinisation; Elderly productivity

**Abstract**

The elderly often struggle when interacting with technologies. This is because the software and hardware components of the technologies are not familiar to the elderly's mental model. This is a lack of empirical studies about how the concept of familiarity can be infused into the design of interactive technology systems to bridge the digital divide preventing today's elderly people from actively engaging with such technologies. In this paper, a multi pronged approach is utilized. We investigate the Effects of Familiarity in Design on the Adoption of Wellness Games by the Elderly, familiarity in productive ageing, familiarity in efficient collaborative crowdsourcing, productive ageing through familiarity based Intelligent Personalized Crowdsourcing and familiarity based Agent Augmented Inter-generational Crowdsourcing. The results show that familiarity in design improves the perceived satisfaction and adoption likelihood significantly among the elderly users. These results can potentially benefit intelligent interface agent design when such agents need to interact with elderly users. A Crowdsourcing algorithm, CrowdAsm is developed. By using CrowdAsm we are able to dynamically assemble teams of workers considering the budgets, the availability of workers with the required skills and their track records to complete crowdsourcing tasks requiring collaboration among workers with heterogeneous skills. Theoretical analysis has shown that CrowdAsm can achieve close to optimal profit for a collaborative crowdsourcing system if workers follow the recommendations.

# List of figures

# Chapter 1 – Introduction

The terms 'personalization' and 'familiarity' have widely been used and researched in various fields. For instance, relevant outcomes in the consumer behavioral research and trust development for the e-commerce websites (Gulati and Sytch, 2008) have been the direct impact of familiarity and personalization. The applications of familiarity and personalization could have important impacts for the creation of a smart city while adequately attending to elderly care.

## 1.1 Research Problem

This research problem focuses on two key challenges of an ageing population. First, it focuses on the personal wellbeing – cognitive and physical – of an ageing population. For addressing the two-dimensional research problem, the role of gerontechnology has been investigated in the context of "personalized familiarity", which shall be elaborated in the later chapters of this thesis. Gerontechnology is seen as a means of bringing positive change in the elderly's lives. This is primarily because gerontechnology in general and health based games in particular, have been reported to bring improvements in the reflexes and core tone of such population, along with proving to be a great cardiovascular workout (O'Loughlin, 2015). Secondly, the paper focuses on Productive Aging with special attention to work-life balance in relation to intergenerational crowdsourcing. The idea has been taken up for an increased involvement of the elderly population in productive activities (Straub, 2014).

### 1.1.1  Personalised familiarity and Older Adults

As people start moving into the later stages of their lives – become elderly i.e. between 50 and 75 years of age – certain physical and cognitive changes occur within their personal spheres. A degree of reluctance to adapting the various aspects of everyday life becomes difficult for such people. This is primarily because they lack the familiarity with that particular environment, be it a general activity or a specific customized environment. Hence, it becomes vital to develop a certain degree of familiarity among these individuals so as to maximize their functional abilities in that very domain (Son, Therrien & Whall, 2002). In particular, older adults with dementia face major difficulties pertaining to familiarity, which limits their functional abilities to a great extent. This implies that there exists a close association between the physical and cognitive abilities of the elderly population and their familiarity with a particular environment. A more common condition underlying familiarity and the elderly is Mild Cognitive Impairment, which is a precursor of dementia (Werner, Heinik, & Kitai, 2013), and thus by instilling personalized familiarity the functional and cognitive abilities of older adults can be fortified. To be specific, technology is somewhat alien to the elderly due to lack of engagement, and thus studies have shown that these technologies can be customized to support their social engagement and wellbeing (Leonardi, et al., 2008).

### 1.1.2 Cognitive Decline

Cognitive decline and ageing have long been associated, and many have investigated the topic. Also commonly referred to as age-associated cognitive decline, this is a condition where individuals start to gradually lose their cognitive abilities as they age; however, this decline varies across individuals and there is no universal marker (Deary, et al., 2009). This decline can be characterized by a variety of components ranging from processing speed, memory, and reasoning to executive functions underpinned as a decline in general cognitive abilities. The severity of this cognitive decline depends on various factors such as lack of engagement and interest, and may lead to acute dementia as a consequence. This decline also has milder conditions like aMCI 'Amnestic Mild Cognitive Impairment' and Alzheimer's disease, which according to studies may contribute to gaps in social cognition and ability to transition from basic to complex situations. Although considered an eventual outcome of healthy ageing, aMCI is considered as a source of decline in episodic memory (Koen & Yonelinas, 2014). This cognitive decline is in turn associated with the lack in familiarity of older adults with the various environments as they age, and thus personalized familiarity is proposed to be the most viable solution to bring knowledge-based familiarity among the elderly population.

### 1.2 Research Question

It has been commonly observed that as people age, they start to lose their cognitive capabilities, which influence their physical abilities as well; a reason why most elderly population representatives are found to be technophobic. The elderly tend to be technophobic. Thus, the research question here is: How to alleviate technophobia and the gap between engagement and familiarity of the older population? The research question also includes how to make the elderly more employable and how to utilize their skills in the form of a meaningful contribution towards solving complex issues that the youth cannot solve by themselves.

### 1.3 Potential Solution

As mentioned earlier, cognitive decline is a common condition associated with healthy ageing, and does not relate to any particular disease; all elderly people experience a decline in their cognitive capabilities – which translates into limiting their physical abilities – however the degree of severity varies across individuals. This variation occurs due to several factors that have already been identified earlier. Therefore, considering the gravity of situation here, many researchers and investigators contend that personalized familiarity is among the best solutions to maintain cognitive and physical wellbeing of the elderly (Leonardi, et al, 2008; Giovanetti et al, 2006). Moreover, since elderly people with AD or dementia have impaired explicit memory but preserved implicit memory, it is proposed that this implicit memory can be utilized by virtue of a developing a sense of familiarity (Son, Therrien, & Whall, 2002). Finally, in order to alleviate the digital divide – the technophobia that exists among the elderly population – it is proposed that gamification be used as a potential intervention (Cugelman, 2013). By inducing pro-technology behaviors through gamification, the pertaining technophobia can be reduced to a great extent. Moreover, as the elderly develop more familiarity, and adapt behaviors that encourage the use of products, the digital divide would be eliminated and a higher usability will be attainable (Boger et al, 2013). Finally, increased trends of intergenerational crowdsourcing – such as Grandparent-grandchild GP-GC relationship – is also proposed to bring a fairly positive change in the cognitive and physical wellbeing of the elderly population (Nagai et al, 2013).

## 1.4 Overview of Thesis

The purpose of this thesis is to investigate the association between familiarity and wellbeing of the elderly population; with particular focus on personalized familiarity and crowdsourcing. Ageing and cognitive decline are two interlinked concepts, that have been the center of academic research for many years. As people age, their wellbeing, characterized by cognitive and physical capabilities, starts to decline and if corrective interventions are not adopted promptly, the situation may lead to drastic outcomes. In order to improve and maintain elderly population's wellbeing, several interventions are proposed that relate primarily to an increased familiarity to various environments. In particular, personalized / technological familiarity and gamification are proposed as key solutions to this problem and to make use of the preserved implicit memory in older adults. Also because a lack of engagement has also been associated with reduced capabilities, increased intergenerational crowdsourcing is also recommended to keep the elderly engaged and increase the usability of various products and environments by the elderly population.

# Chapter 2: Personalised familiarity

## 2.1 Familiarity Theories

The terms 'personalization' and 'familiarity' has widely been used and researched in various fields. For instance, relevant outcomes in the consumer behavioral research and trust development for the e-commerce websites (Gulati and Sytch, 2008) have been the direct impact of familiarity and personalization.

Similarly, 'personalization' and 'familiarity' has various applications in the field of gerontology. Along with the recollection process, familiarity tends to improve the recognizing ability by accessing the information present in brain, already (Bressler and Ding, 2006). Many previous studies have shown that the aging process damages recollection feature, but has significantly little or no effect on the familiarity. Thus, familiarity can be implemented in the recognition memory betterment of the elderly people.

In 2000, Pinto et al. emphasized that gerontechnology along with ergonomic approach can result into improvement of life quality and activities. The study put forward some suggestions in the designing of home entrance and kitchen, which could result into increased level of independence and safety for the old people. The main finding was that the homes, where old people have spent most of their lifetime, are easily adaptable for them. This not only conserves their lifestyle but also increases their confidence level. Again in the same year, 2000, Pinto et al. demonstrated the impact of the same approach on the elderly patients suffering from cardiovascular disease. The level of satisfaction was remarkably increased in them by boosting the sense of responsibility and motivation. Thus, gerontechnology sets as a means of upbringing dynamic health related changes in the old-declining patients.

In 2002, Son et al. proposed that familiar environment can serve to increase the functional capacities in the elderly people suffering from dementia. A proposed model was constructed by using literature from different sources, through clinical observations and previous experiences. It was found that the people suffering from dementia have preserved implicit memory. This can be utilized along with the familiarity, to bring about positive changes in the patients.

In 2004, Demirbilek and Demirkan studied the impact of the environment on the ageing factor of an elder person. Mostly, the people prefer to live in their familiar environments in the 'getting older' phase of their life. This sets as a prime important aspect to design homes for the old people, and increases its safety, usability and attractiveness. The study suggests methods such as brainstorming, scenario building, unstructured interviews, sketching, and videotaping techniques to be used in order to increase the participation from the patient. Moreover, quality deployment matrices should be used to analyze relationship between the patients' requirements and design specifications. The studies emphasized on the participation of the elders in designing the house for them, as this enhances the design solutions, as well as increase the satisfaction level in the subject.

In 2005, Sung and Chang conducted a review to study the impact of music on the old people, and showed that such interventions can have positive effects in the patients' life. Significant results were found in the people suffering from dementia problem. The patients exhibited reduction of agitated behaviors, as an outcome of their favorite music.

In 2006, Giovannetti et al. implemented the advantages of using familiar objects for various dementia patients. Various tests such as Naming, Gesture, Semantic/Script

Generation, and Personal Object Decision were used to analyze the familiarity phenomena in the patients. The results showed drastic statistical significance in the personal and analog objects. The familiar objects exhibited better gestures and information details. The study encouraged these results to be implemented on the dementia patients. The dementia patients should be provided with the familiar objects for their better use. Furthermore, the dementia patients can take benefit of new objects, with increased pre-exposure and familiarity development to them.

In 2005, Mitchell et al. showed that the familiarity can improve interaction ability. With the help of an interactive software application, the improvement in responding of the students was affirmed. In 2005, Musha et al. formulated a therapy that involved the use of a robotic pet. This study proved that implementation of such personalization therapy can cause neuron activation of the brain cortical for the patients getting older. This therapy was given for 20 minutes along with 2-hour art therapy, and it brought about significant improving results in the patients under observation.

In 2006, Turner and van de Walle described familiarity in HCI approach. They suggested that the familiarity is applied by simply "coping with situations, tools, and objects, or more specifically by understanding the referential whole.". The younger generation has intimacy with the technology and interactive products, but the ageing adults do not get engaged with the technology. It is supposed that rather they seek that how the technology "meet the demands, hopes and aspirations of their everyday life". Turner and his colleagues have focused on familiarizing the elderly people with the prevailing information technology. For this, they interviewed them about their computer learning and its effect on their lives.

In 2008, Leonardi et al. proposed a design to convert the technology language to a familiar one, which could easily be understood and adopted by the elderly people. The interface was supposed to be interactive by physical and cultural means. This approach ruled out the traditional windows.

In 2008, Turner conducted an empirical study on the familiarity, which involved groups striving to learn about the computer and its services. The study suggested the incorporation of technology in the everyday life, to help in its understanding. Moreover, the daily practices need to be changed in accordance to the requirement, in order to get better comprehension of the technologies, apart from designing for easy use and experience. The technology is meant not only to make practices efficient but it is responsible to change them entirely. Therefore, research is required to explore the technological horizons in terms of the evolving practices.

In 2008, Barry encouraged the incorporation of familiarity into the home design plans, by the designers in order to bring positive changes in the old people. The houses of elderly people should be sensitive towards their habits and routines. Due to undesirable time changes, the older people might not be able to continue living in the same house, to which they usually adopts familiarity. They might have some sort of feelings, sense of identity and attachment associated with their house. However, they still seek for comfort level in the new locality through previous habits. Therefore, the design of a home, where an old person has to live, should tend to preserve all her character attributes, which defines him. The study also formulated an ideal home, with the incorporation of the habits and routines of an old man.

In 2010, Algarabel et al. studied familiarity in the patients of Parkinson's disease and Lewy bodies' disease. The extent of familiarity in the patients was estimated by using recognition for word targets and distractors. The results showed that the patients of Parkinson's disease and dementia can use familiarity process, despite of the diseases' impact on their memories. Though, in the late stages of these diseases, the familiarity phenomenon is highly affected.

In 2010, Weiermann et al. studied the relation of Parkinson's disease to the memory process. Both the recollection and familiarity were analyzed, which may be effected because of this disease. The comparison between the Parkinson disease patients and healthy old people showed that the recollection remains intact, but the familiarity process is selectively affected.

In 2011, Jurjanz et al. studied the impact of amnestic mild cognitive impairment on the familiarity. They show memory dysfunction and deficits in contextual knowledge. This leads to problems in the identification of objects and person around them. The relation of the neural networks was investigated within such people with regard to the familiarity. The results depicted decreased activity in the right prefrontal brain regions, with regard to familiarity. This can result in reduced social cognition, and task management in the patients.

In 2011, Chrysikou et al. envisaged a relation between the semantic dementia (SD) and the familiarity loss. The study focused on the effect of SD on the loss of conceptual knowledge in the patient and investigated the autobiographical experiences to devise methods of maintaining the object concepts in the patients. By using tasks such as naming, gesture generation, and autobiographical knowledge for familiar or similar objects, the study was conducted. The results showed that significant dissociation relation between performance and assessments exists.

In 2013, Silveira et al. proposed an increase in the activity of the old aged patients through an IT-based system. This proactive training application was especially made for the people who cannot exercise due to their growing age. Moreover, it was meant to assist the patients to follow training plans along with increase in their social activity. This application proved to be highly motivational for the patients by helping them in monitoring their performance, and hence is a source of encouragement for similar mobile-based incentives.

In 2013, Boger et al. suggested utilizing the products of higher familiarity to bring betterment in the condition of the dementia patient. This approach can enable the patient to move independently, without the need of a caretaker. The study assessed the impact of familiarity on the use of different faucet designs by dementia patients. The findings suggested that familiarity could turn into lower assistance level, lesser operational problems, and higher satisfaction levels. Such a prospective can be implemented to various other products and tasks, in future, to increase their usability.

In 2013, Werner et al. conducted a study regarding the familiarity, knowledge, and treatment options for the dementia disease. The findings emphasized the knowledge of physicians and their preferences. The research serves to provide guidance about the dementia and its symptoms among the physicians.

In 2013, Jaffer et al. suggested the use of Internet in the medical monitoring by the trainees. The Internet serves to shorten the distances and increase the interactivity across the distances. This can be used in the medical perspectives, to bring about positive changes. Such a technology is Web 2.0 and 3.0. A workshop was arranged to bring about the task along with the questionnaire section. The study encouraged the virtual mentoring system and opinion setting in the medical staff, according to the overall questionnaire results.

In 2013, Cugelman provided a view on the gamification and its effects on the health behavior change. He described the principles, mechanisms, and evidences governing these changes and their efficacy in the patients' health betterment. Moreover, the principles for both the gamification and the resulting behavioral change were interrelated. A criteria of assessing the gamification strategy was also provided to be implemented in the digital health interventions. The study encouraged the engaging use of gamification in the digital interventions by finding more innovative ways in future and highly socializing the approach.

In 2014, Koen and Yonelinas conducted a review to formulate the relation between the diseases like amnestic Mild Cognitive Impairment (aMCI), and Alzheimer's disease (AD) and the memory decline. The effect of these diseases was checked on the two important aspects of memory; recollection and familiarity. The results showed that the aging has direct effect on the recollection phenomena, whereas, the familiarity aspect mostly remains intact. The diseases aMCI was also found related to the decline of recollection. On the other hand , the AD was associated with both the recollection and the familiarity. The familiarity decline was also found to act as a behavioral marker for the patients that might develop dementia in the near future. The study further suggested that the recollection is critically linked to hippocampus, whereas, the familiarity is associated with the perihinal cortex region.

## 2.1.1 Familiarity and personalized familiarity

In order to understand personalized familiarity, it is important to break the term down and consider their individual definitions. 'Personalized' means adjusting or aligning components / features of a particular environment, situation, area, or product / service according to the requirements of a person. Familiarity on the other hand is defined as having the knowledge or mastery of something (Dictionary.com, 2015). Thus, personalized familiarity can be defined as aligning or customizing the attributes and features of a particular place or thing in order to enable an individual develop relevant knowledge or mastery.

Familiarity has many different types that can generally be characterized as familiarity with a person, place or thing. To be more specific familiarity may range from familiarity with technology, familiarity with a particular area or environment such as the kitchen or the living room, familiarity with the neighborhood, and familiarity with processes to familiarity with food and health characteristics etc. Although the aspects of familiarity vary to a great extent, this thesis is only going to focus on a few most relevant aspects. These aspects include personalized familiarity with interactive and personalized games and familiarity with intergenerational crowdsourcing. This is done with the help of the portrayal of some highly interactive games and other activities.

## 2.1.2 Personalised familiarity and Recognition Memory

Recognition and familiarity are closely associated with each other. Recognition memory and its deficit are often considered to be associated with Parkinson's disease. It has been contended that the patients that have acquired this disease have deficit in recognition because of the low retrieval requirements of the task and not the encoding deficits (Algarabel, et al, 2010). However, there exist several contradictions pertaining the association of Parkinson's disease with impaired recognition memory (Weiermann, et al, 2010). Nevertheless, regardless of these contradictions, recognition memory and familiarity have been reported to have some degree of association, especially in elderly population. Those representatives of the elderly population who are less familiar with a particular phenomena may be expected to experience a greater lack of recognition memory. This implies that unless an individual is familiar with a particular concept or product or thing, he or she cannot be expected to recognize it for future references. As Yonelinas (2002) elaborates, familiarity means that the individual feels that a similar event has happened before, without recalling the particular encoding text, it becomes easier to bridge the association between personalized familiarity and recognition memory. By inducing a higher degree of personalized familiarity among elderly populations, recognition memory can be be improved to a greater extent.

### 2.1.3 Personalised familiarity and Gerontechnology

Similarly, the 'personalization' and 'familiarity' has various applications in the field of gerontology. Along with the recollection process, familiarity tends to improve the recognizing ability by accessing the information present in brain, already (Bressler and Ding, 2006). Many previous studies have shown that the aging process damages recollection feature, but has significantly little or no effect on the familiarity. Thus, familiarity can be implemented in the recognition memory betterment of the elderly people.

*2.1.4 Encouraged use of IT in the medical field*

The medical and healthcare paradigm has spearheaded over the years – the past 2 decades especially – by virtue of an increased utilization of Information Technology. From treatment to prevention, and from efficiency to effectiveness, use of Information Technology has revolutionized the realm of medical sciences. IT has especially played an integral role in supporting the wellbeing and lifestyles of those who experience some type of impairment or inability. For example, the development and utilization of accessibility devices, prosthetics, and devices equipped with artificial intelligence in a variety of ways has nurtured medical sciences significantly. Key areas where IT has played a profound role include biotechnology, pharmaceuticals, medical device and equipment development, and other innovations. A recognized example of the use IT in medical field is EMR or Electronic Record Management, which has enabled healthcare providers and associated professionals improve the quality and correctness of services being provided to patients. More pertinent examples – relevant to the current thesis of personalized familiarity and wellbeing – include the utilization of smartphones, tablets, tele-health services, and other key mobile technologies. These technologies are making it easier to overcome the digital divide that exists among the elderly people who either suffer from dementia or cognitive decline of some sort.

## 2.1.4 Definition of Personalised familiarity

As elaborated earlier, in order to gain an understanding of the term 'Personalized Familiarity', it is important to break the term down and consider their individual definitions. 'Personalized' means adjusting or aligning components / features of a particular environment, situation, area, or product / service according to the requirements of a person. Familiarity on the other hand is defined as having the knowledge or mastery of something. Thus, personalized familiarity can be defined as aligning or customizing the attributes and features of a particular place or thing in order to enable an individual develop relevant knowledge or mastery

### 2.1.5 Comparisons of existing frameworks and choice of a Framework for Personalised Familiarity

The effect of familiarity on the interaction between a user and a technological product can be summarized in three stages. The first stage relates to the works of Fitts and Posner (Fitts and Posner, 2010), Taatgen et al (Taatgen et al., 2008), and Ericsson (Ericsson and Towne, 2010), and can be referred to as the "Cognitive stage". It revolves around the basic knowledge required for task execution. The second stage is referred to as the "Associative stage" in which the persons move from their declared understanding of the action to a process based performance of the action. The stronger the association with something the person is already familiar with, the higher the speed of execution and the lower the amount of effort needed to carry out the action (Taatgen et al., 2008). The third stage is called the "Autonomous stage" in which the actions follow a smooth procedure and actions are quick and involve negligible effort.

While innovation on a fundamental level could repay an age's portion related changes in their physical, social and intellectual assets, in this manner improving their personal satisfaction, the elderly much of the time get impartial and baffled by excessively complex innovation. Other than procedures to persuade and prepare more seasoned clients to utilize innovation, there is additionally a requirement for gadgets that are better customized to the abilities of a maturing client. The ordinary maturing procedure is ordinarily accompanied by visual and sound-related hindrances, and a decrease in working memory, particular consideration, and engine control is examined.

Based on this three-stage model of the effect of familiarity, we need to propose a framework of familiarity design for human computer interactions. The following are the proposed frameworks for the elderly.

*2.1.5.1 Netcarity*

NETCARITY aims at such artifacts, in order to make them easily and immediately understandable by the elderly. This is due to the syntax and the semantic, which are derived from the experience of the elderly people. NETCARITY is derived from the following process design:

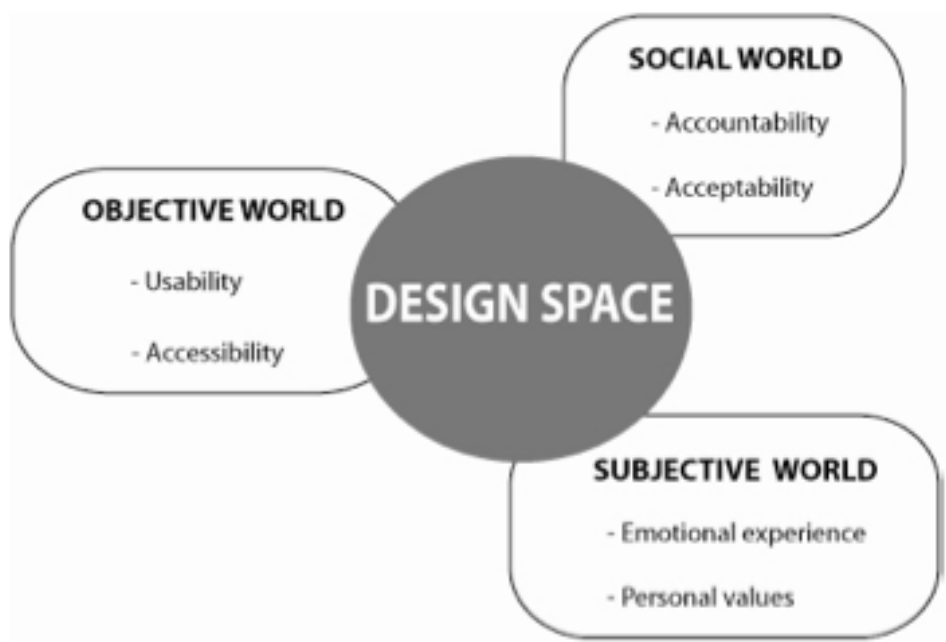

**Figure 1: Netcarity**

However, the drawback or weakness of this system is that the requirements of the

NETCARITY are not suitable for the interface of the windows, icons, menus and pointing (Leonardi, Mennecozzi, Not, Pianesi & Zancanaro, 2008). The interface of the WIMP is largely used with the help of a mouse and a keyboard but the NETCARITY involves actions like drag, point and click. Moreover, the secondary elements are used to obtain the desired effects that might again be a little difficult for the elderly people. On the other hand the NETCARITY has an interface that involves the concept of plainness. The decorative elements in the visuals are avoided in the program. These decorative elements do not serve any suitable goals thus they have been removed. A simple design has been adopted for the NETCARITY (Leonardi, Mennecozzi, Not, Pianesi & Zancanaro, 2008). Secondly, there has been a usage of strong contrast for the elderly people so that differentiating between objects becomes easy for them and so that they do not get confused. Moreover, simple language can be used in the text messages.

Opacity layers have also been introduced in the framework so as to differentiate clearly between those objects that are active and those objects that are passive. The digital objects are graphically represented in a way that is stylized and not much realistic (Leonardi, Mennecozzi, Not, Pianesi & Zancanaro, 2008). Clutter has also been avoided in the program and the animations are kept smooth, as the quick ones are not perceived in a proper manner by the elders.

The following diagram shows how a user-friendly interface can be used for the elderly people:

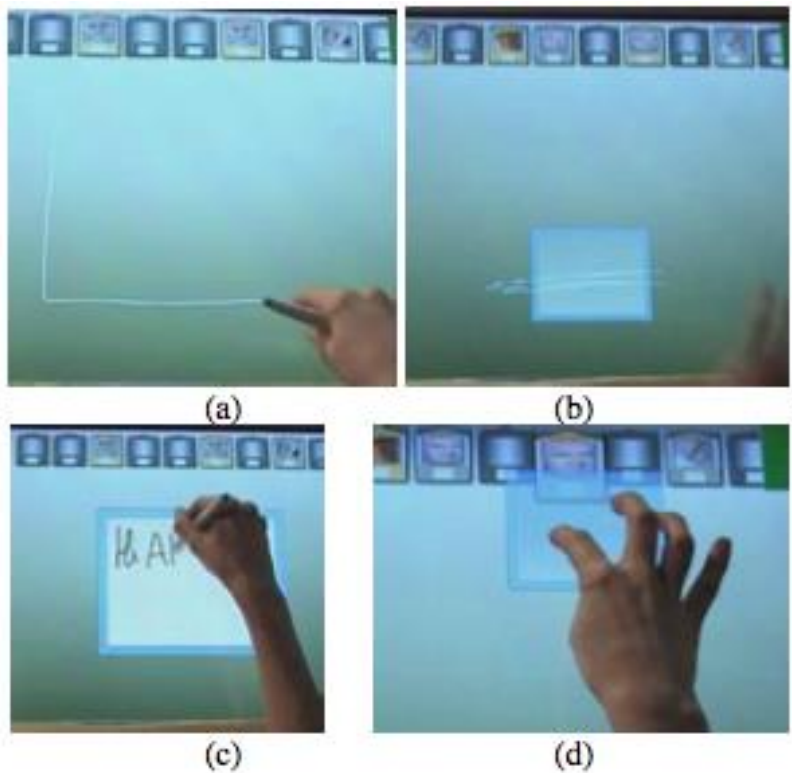

**Figure 2: User-friendly interface for the elderly**

In the diagram the solutions are based on the elderly familiarity process. The first diagram shows how a new empty document can be created with the usage of gestures. The second diagram shows a gesture for deleting the document. The third diagram shows the process of writing with the help of a pen. The last diagram shows the gesture of sending a friend a message.

This is an acceptable interface for the elderly people as gestures are involved in the process, which are easier to use for the elderly. However, it might be difficult for the elderly to remember so many gestures all at once. For example creating a new document would involve the making of an L shape and it is no way related to creating a new document and might result in the elderly person forgetting the gestures. Such gestures should be developed so that the elderly person is able to remember these gestures easily. Moreover, this method is further limited, as it cannot be used on desktops and windows.

*2.1.5.2. Privacy framework*

As the modern world of technology is advancing with time to meet the needs of the age population, the technology designers are thinking of producing a gerentechnology field and modifying it with time reconsidering the real life problems of the older adults. (Achenbaum, 2010). In these circumstances, as the people are concerned about their privacy limits, the swift evolution of social networking is occurring by bringing the privacy changes.

The findings from the works of various researchers suggest that adults are not concerned about their privacy, but since concepts of date mining and aggregation are unfamiliar to them, they like this framework, because they are mainly concerned about hiding their financial matters from their children.

Following are the domains of this framework:

- The right to be left alone (Wareen and Brandies, 1890)
- The right to self determination (Introna, 2003)
- The right to determine the uses of personal data (Bloustein, 1968)
- The right to determine physical and virtual boundaries
- Data is transparent and verifiable. (Odlyzko, 2004)

Following are the technology features used by this framework:

Internet-enabled medication dispenser, which alerts if a medicine is not taken on time

1. Games to promote health

2. Clock with sensors and lights to facilitate awareness between remote family members.

3. Wild Divine: Biofeedback game aimed at reducing stress (commercial product)

4. Mirror Motive: mirror that displays reminders (e.g. medicines) and coordinates social engagement when person is detected nearby

The technology devices designed according to this framework were home based rather than big devices. They were very normal and considered one of the daily routine things. However, assistance is still given to the users so that they can easily handle it, and the concept of familiarity can be handled. The products are very easy to use and common.

With the integration of this framework into the daily routines of the elderly community, their life will become much easier, and they won't need any further care giving and they can be independent all over again.

The only drawbacks of the products used here as stated by some of the participants in the experiments are constant interruption of noises in their private lives. Moreover there shouldn't be excess of priority given to technology rather than human contact. This development can be constructively informed by trans disciplinary theories of aging, which link gerontology to real-life issues (Achenbaum 2010).

### 2.1.5.3 The use of I.T devices for active ageing

The UTAUT model worked on the intentions of elderly people to use smart phones for e-health services. There are several factors that do not contribute to the usage of a smart phone completely. The reasons for this are that elderly people either find smart phones very uninteresting or they find them difficult to use. But according to this model, the usage of a smart phone itself can offer an elderly person in taking e-health services in many aspects. (Boontarig et al, 2012) However, the Graphics User Interface that is provided as an application and many other of the sort on a smart phone need to be simple If the elderly could understand smart phone interface for e-Health services, it would be help reduce effort and difficulty to use. Hence, this should be considered as an influencing factor for the using a smart phone for e-Health services. The limitation of this study is the focus on specific characteristics of the respondents with good education and knowledge in technology.

Similar can be the case with a computer. There can be a lot of uses but there is a lack of familiarity. (Alm, 2002). The technology designers should think of producing much more appropriate and easy to use devices that can help the elderly people in their daily lives.

### 2.1.5.4. The Assistive Technology (AT) MODEL

The AT model stands for assistive technology, and it is developed for assisting the elderly people with their daily house chores and make them independent. (McGreadie & Tinker, 2005) The benefit of this model is that with its products, it allows an elderly individual to perform those tasks that they were otherwise physically unable to do. It ensures a high rate of safety and reliability. This is not to say that technology should take the place of humans in case of everyday care and being handy and home, it generally focuses on the daily routine and the physical and mental chores that elderly people are unable to do, and improve their quality of life with robots and other technology.

Elderly people who are usually disabled need a lot of special care and respect. For example they need to grab rails or lifts instead of stairs. There is plenty of need for social alarms, there can also be arrangements made of reading text messages on the TV. Special Braille readapts can also be introduced.

There are a lot of elderly people who do not like introducing AT to their lives; some even couldn't get easy with using them because they were unable to. But every person makes adjustments accordingly, depending on their social needs and boundaries. Paradoxical situations could be found where they did not feel comfortable with the alert alarms. Many people preferred their individuality or privacy at homes.

*2.1.5.5 Conclusion and Selection of Framework*

Human life has greatly evolved with the introduction of technology and its various uses in different aspects of society. It has even helped the elderly community in not only fixing their many diseases and impairments, but also provides them a company in their loneliness. Various approaches and models have been suggested with positive and negative points in each of them. However, the framework that appeals the most to the lifestyle of elderly people is the personalized privacy framework. The benefit of this framework is that the products it offers helps in various aspects and they are very easy to handle just like home made things. Also, adults feel comfortable with it due to the fact that they can stay independent and active as well.

## 2.2 Potential of crowd sourcing

Crowd sourcing is a technique used to conduct a public participation for transit planning (Brabham, 2010). It is an online and problem solving approach and can propose solutions and ideas for the whole population in an easy manner. It is a public involving method for government administration. And if in this aspect, we say that such a considerable proportion of our population is neglected just because they are not familiar with the modern technology and its terminology would be absurd. This basically provides the major source for their exclusion and isolation from the society because the technology introduces them to a series of elements and terms, which they never came across before, and they feel all alien and technophobic. (Yu, Shen, Miao, and An, 2012).

There must be some programs that help them in updating themselves about the new software and hardware, so that they can not only stay in touch with their families and friends, but help the world in giving their opinions too. We all need to learn from history and get guidance from our old age community, because they are more experienced and rational thinkers. For example, in order to conduct a survey on EEG brain data for the study, people ranging from 18 to 89 contributed. This helped in bringing more authentic results. (Kovacevic, 2015) Their contribution and participation does count in surveys, research and opinion polls. So for this reason, there should be respective ICT programs, which impart knowledge to the elderly about technology and help them in using it and staying updated.

Crowd sourcing shortly termed as CS, offers dividing large problems into smaller ones by distributing them over a diverse area and solving those things that are impossible to be done by the computer. (Yu, Shen, Miao, and An, 2012). Crowd sourcing is a very feasible approach for getting and giving help in a lot of domains in a society. For example in order to acquire data pertaining to the public, we can use the GPS methodology and collect information. But definitely, the views and opinions of a 30-year-old person do vary from those of a 60-80 year old, and we certainly cannot neglect the latter age domain. So in order to have a valid public opinion and letting the public be the local scientists, we need to introduce and promote greater use of technology by the elderly. Utilization of this online source is a very promising avenue for data collection and problem solving. It can catalyze the solicitation of ideas and solutions from people and from various aspects of the same communities (Brabham, 2010).

According to Orlov (2011), $21^{st}$ century is an era in which men and women aged 85+ represent the fastest growing demographic; and service providers, and product makers need to understand enough about how our increasingly senior society thinks and wants to interact; what are their needs and what they do not like. So for this reason, a technology mainly named as the linkage survey technology, caters to the needs of the senior class of the society and collects their opinions of things, and their familiarity with the technology. It is a Mason and Ohio organization, whose priority is to meet the needs and resources for a community that is ageing, made surveying this population feasible by finding a broad response across all senior community.

The elders in general go through tough times when they want to interact with and learn new technologies. There should be such designs, which are user friendly for them and make them get over their technophobia (and they can use the technology as much as they can in order to carry their works out themselves and also play their roles for the betterment of society wherever necessary. (Leonardi, Mennecozzi, Not, Pianesi and Zancanaro, 2008)

This research and the efforts for the inclusion of the elderly under the umbrella of technological use by the population have important ramifications for society. A greater embrace of technology by the old age community would lead to provision of greater support to them and their being independent. It will also boost their morale and level of encouragement, and they will feel good about themselves that they are responsible and that they did participate in the well being of society. This will also improve their level of motivation, quality of life, and standard of thinking. So in order to make them comfortable and to cherish their opinions, such language and atmosphere should be designed where they can comprehend and feel at ease. (Brabham, 2010).

Our generation has grown up in an era of technology, where we can comprehend and know how to interact with technology. All this has created a common ground of knowledge and language that we can understand and all the practices are instilled in our habits and that is the reason we don't think technology to be such a big issue or a phobia (Leonardi, Mennecozzi, Not, Pianesi and Zancanaro, 2008) We all admit the fact that in order to meet the needs of modern society, we need to know and have all the skills of technology usage, be it academic, social or pertaining to our own relations and personal needs. In such circumstances, we cannot just sit and do nothing about the elderly community not being comfortable with it. (Leonardi, Mennecozzi, Not, Pianesi and Zancanaro, 2008). During interaction with technology, it is common for the elderly to give up and reject it. This is not the solution to the goodwill of a prosperous and developing nation. According to WHO (World Health Organization), it is necessary that there are such policies and programs which can keep the older community active and helping, so as to promote the sustainability of the prevalent society. Keeping an active aging is one of the biggest challenges our society will have to face in the coming decades, and we will find it difficult to promote equal opportunities for all. (Judice et al, 2010). The generation before us, or let's say our elder community have not lived in an era with such circumstances, they lack that ground of experience and they do not have the proper knowledge; nor can they relate with the new technology. The trend is changing and for all this, a common ground of information or a common language has to be worked upon, which can bridge the divide preventing today's elderly age people from interacting with technology (Pan, Miao, Yu, Leung and Cui, 2015).

The combination of sensory and cognitive changes as well as the loss of social and economic status apparent everywhere can give us a hard time in making this variety of population in absorbing and up taking the new technology (Isaacs, Martinez, Scott-Brown, Milne, Evans, and Gilmour, 2013).

### 2.2.1 Crowdsourcing, familiarity and elderly speech

There have been certain speech soft wares used which are used for input of information; the common among them being the ASR (Automatic Speech Recognition), but even it is not efficient or sufficient enough to meet the needs due to the reason that human speech changes with time, and it is difficult for the ASR interface to transcribe all of the waveforms. Recognizing speech becomes even a more difficult problem in other languages, because of the lack of the speech training data (Judice et al, 2010). For the elderly speech data collection for the purpose of crowd sourcing, Doarazov can be used, which can help in collecting data. But apart from all this, special care is still not taken of the font size, having way too many instructions on the front page, and the sessions being too long. This was what made the elderly people give up or get discouraged and didn't provide a user-friendly atmosphere to them at all. However, the problems have been worked upon and solved up to a greater extent by magnifying the font and other. Bringing familiarity in Crowd sourcing and other technology is of vital importance. What the designers usually take basic care of is the 'look and feel' process of an artifact. (Judice et al, 2010). How the interface should meet the sensorial abilities of the users is of main concern to them. Using contrasting colors and section for the sake of avoiding confusion as well as simple language rather than that of the technical terminology provides a very natural background for them. The animation, opaqueness and the whole graphic styles are also worked on so that the level of familiarity can be retained. Modern world can prosper with this.

### 2.2.2 Intergenerational crowd sourcing

While the merits of crowdsourcing have been described earlier, intergenerational crowd sourcing is another powerful idea. While crowdsourcing is a useful and powerful tool, sometimes the pool that makes the "crowd", lacks the technical expertise or the relevant experience in order to accomplish the tasks. Here, people from different generations can collaborate to solve the problem. For example in solving problems where the youth does not have linguistic expertise, while the elderly know more dialects/languages or in instances where it would be too expensive to hire the youth while the elderly would happily complete the same job at a lower rate, for example in proof reading tasks (Kobayashi et al., 2013). Kobayashi et al. have introduced a collaborative crowdsourcing model that attempts to make maximum use of the elderly's linguistic skills by encouraging the youth to support the elders in overcoming their lack of technical skills.

### *2.2.2.1 Intergenerational crowd sourcing and older adults*

Ageing is an evident fact that has prevailed in this world since the beginning of mankind. People age with time, and as they age, there occur several mental a and physical changes. Being an older adult can be hard at times, especially in the contemporary society that is technology driven. This basically due to the reason that older adults are comparatively less adaptive to the changing dynamics compared to the younger population. Hence, a significant body of research exists that has intended to explore what it means to be an older adult, and what are the key challenges that come along.

Older adults are defined keeping in mind several important factors including their chronological age, transitions in their functional abilities, and changes in their social involvement / engagement levels. In developed economies, the older adults are defined as those who have received retirement from their service and are around 60 to 65 years in age (WCPT, 2015). There are other sources that characterize older adults as people aged 55 years and older. However, there is no universal definition of older adults, but there sure are similarities among these definitions. For instance, from a researcher's point of view, an older adult is characterized as a person aged between 65 to 75 years of age.

### 2.2.2.2 Effects of Ageing in older adults

In order to completely understand the benefits of inter generational crowd sourcing, we need to understand the issues that arise as a result of ageing and how inter generational crowd sourcing can still make good use of the elderly's skills despite their advanced age.

Aging is a natural phenomena that can – under no circumstance – be avoided or prevented. As the time passes by, the human body changes in terms of cognition, physiology and psychology. While some changes / effects of ageing – physiological – may be obvious, there are others that are more subtle. These effects of ageing vary significantly from person to person, and these variations depend on the facilities available, the life experiences, and the environmental exposure along with some others. Where one older adult may lead a comfortably healthy life, remain alert and proactive, another older adult may be experiencing completely opposing consequences. Among the common effects of ageing include health conditions like osteoporosis, osteoarthritis, and some cardiovascular diseases. But more relevant effects of ageing include a decline in the cognitive abilities, memory loss, dementia, AD, and Parkinson's Disease, which in turn lead to decreased physical capabilities of older adults as they continue to age. This reduction in cognitive and physical capabilities results in a reduced quality of life, creating greater dependence on others, and lesser degree of familiarity with the dynamics of the society. Ageing has also been associated with a decreased level of social involvement and interaction, and hence research recommends a greater inclination towards intergenerational crowdsourcing so as to keep the older adults more involved and active.

### 2.2.2.3 Older Adults, familiarity and personalized familiarity

While analyzing the importance of intergenerational crowd sourcing, it is also important to realize the relationship between older adults, familiarity and personalized familiarity.

Now that a thorough research has been conducted, it becomes easier to develop an understanding of older adults and familiarity, and the relative impact of personalized familiarity upon them. As adults age, and move into the later stages of life, they start to experience multiple changes in the mental and physical capacities – usually marked by a decline – which compels them to limit their social involvement, and interaction with other members of the society. This ultimately results in a decreased familiarity to the various situations, since these older adults lack the knowledge of whether such a place or thing exists. Despite knowing the fact that many of these transitions actually enable humans to make their lives easier and more productive, majority of older adults lack the knowledge, and hence end up without any autonomy or independence. To cope with this challenge, personalized familiarity envisions to create knowledge based familiarity keeping under consideration the respective needs and cognitive and physical requirements of the older adult. This results in increasing the quality of life.

### 2.2.2.4 Impact of 'personalization and familiarity on various life aspects of elderly people

**Impact on the confidence level:** Familiarity has a direct relation with the confidence level of older adults. With lesser familiarity about the current trends, older adults lack the confidence to stay involved in the society, whereas, with the help of personalized familiarity, these individuals regain the confidence and consider themselves as an integral component of the society.

**Impact on the health:** Health impacts of personalization and familiarity are an important consideration. With lesser familiarity of the latest advancements in the medical field, the older adults fail to capitalize on all the offerings that can help them better manage their health conditions. With greater personalization, the elderly population gets the liberty to align medical facilities as per their respective requirements.

**Impact on the functional capacities:** Functional capabilities of the elder population may become limited as a result of lesser personalization and familiarity. For instance, if an older adult with motor impairments does not have access to necessary Assistive Technologies, or does not have the knowledge to use them, he would not be able to perform those functions.

**Impact on the satisfaction level:** The satisfaction level is directly associated with the level of fulfillment. As long as the older adults have access to personalization, and have knowledge-based familiarity of the pertaining trends, their satisfaction levels are high and vice versa.

**Impact on the interaction ability:** Older adults' ability to interact also depends on their familiarity with the current societal trends and dynamics. If they lack the knowledge, their interaction ability becomes dormant.

**Impact on the agitated behaviors:** Lesser degree of familiarity with the societal dynamics and lesser access to personalization is directly associated with greater degree of agitated behaviors. This occurs primarily as a result of isolation from the society, and the feeling that their role no longer exists in the society.

**Impact on the use of objects:** Elderly people can only be eligible to use objects as long as they hold the knowledge and information on how to use them, and what benefits can be retrieved from their use. Thus, lesser familiarity and personalization implies a greater limitation in the use of objects.

**Impact on the use of information technology tools:** Information technology tools are complex in nature, and require a greater degree of knowledge and awareness. In particular, smartphones and tablets along with facilities like the Internet call for an equivalent degree of familiarity, and a considerable personalization as the elderly have various conditions.

**Impact on the adoptability of new places:** Adoptability of new places depends on an elderly person's ability to dwell into that place's environment and atmosphere. It also depends on the degree of acceptability that exists within the members of that place towards the new person. Hence, lesser familiarity with the new place naturally leads to a limited adaptability, and would take time.

**Impact on the use of faucet:** familiar faucets correlated with lower levels of assistance from a caregiver, fewer operational errors, and greater levels of operator satisfaction. Aspects such as the ability to control water temperature and flow as well as pleasing aesthetics appeared to positively impact participants' acceptance of a faucet. The dual lever design achieved the best overall usability.

### 2.2.3 User augmented crowdsourcing

The world population is ageing rapidly due to falling fertility and mortality rates. By 2050, the number of older persons will exceed the number of youths for the first time in history (U. Nations, 2002). As the old age support ratios1 for many aging societies start to decline, governments are looking for ways to enable retirees to re-enter the workforce. For example, in Singapore, the WorkPro2 scheme has been proposed to entice retirees to take up part-time or full-time employment while providing incentives to businesses hiring senior citizens.

In the foreseeable future, inter-generational interactions will be an everyday occurrence in the workplace. However, such interactions might not always happen in the physical world (Yu et al., 2010), (Wu et al., 2013). Recent reports from Japan suggest that the elderly are starting to turn to crowdsourcing for jobs.3. Crowdsourcing, which refers to the arrangement in which contributions (e.g., observations, contents, and services) are solicited from a large group of unrelated people (Doan et al., 2010). Some crowdsourcing tasks require workers to perform tasks in given locations (sometimes with the help of mobile devices). This type of crowdsourcing is referred to as mobile crowdsourcing (Yu et al., 2015). Due to the high level of flexibilities, online or mobile crowdsourcing systems present a good match to the elderly's life style, which emphasizes work-life balance.

Although existing commercial crowdsourcing platforms, such as the *Amazon's Mechanical Turk (mTurk)*, mostly rely on workers to pro-actively look for and take up tasks, research works in artificial intelligence (AI) and multi-agent systems (MASs) (Yu et al., 2011) are moving towards building intelligent agents automate the allocation of crowdsourcing tasks in a situation-aware manner (Pan et al., 2009), (Shen et al., (Yu et al., 2010), (Yu et al., 2013), (Yu et al., 2014), (Yu et al., 2015). Nevertheless, existing research in this area has not yet taken inter-generational interactions into account. The subject expertise from the elderly and the technical capabilities from the young can be combined through crowdsourcing to tackle complex and challenging tasks (Kobayashi et al., 2013). However, inter- generational crowdsourcing has also posed an important open research question to the intelligent agent community - what motivate people to join crowdsourcing?

This question is related to the incentive mechanisms research, which is an emerging area in the field of MAS. Gao et al., 2012 studied different designs of crowdsourcing contests for motivating worker participation. In (Araujo, 2013), the contest dynamics on 99design.com are analyzed to understand how financial incentives can be translated into quality outcomes. Jacques et al., 2013 proposed 3 conversion-rate-based metrics to help measure the willingness of workers to participate in paid crowdsourcing tasks. A time-varying payment scheme has been proposed in Difallah et al., 2014 to retain the interest of workers who need to work on long batches of similar tasks. In Naroditsky et al., 2014, the effectiveness of different incentive schemes to encourage workers to join viral referral activities has been studied. In spite of these works, there is a lack of large-scale research studying the factors motivating inter-generational interactions in crowdsourcing.

### 2.2.4 Crowdsourcing, elderly needs and technology for health via gaming

A new study discusses a new genre of specially created video games that provide therapeutic value. In a new publication, researchers from the University of Utah discuss how video games can be used to help patients with cancer, diabetes, asthma, depression, autism and Parkinson's disease.

Technology and video gaming has become a part of personalized medicine. It can also offer several physical and mental therapies. Both will be discussed here.

The increase in age can result in several impaired mobility cases. For many older adults, a lack of independence is also closely associated with a decline in mobility, which may result from a range of medical conditions such as hip fracture, arthritis, stroke or other injury and can itself be exacerbated by physiological aging and inactivity. Mobility impairment can increase the risk of illness. (Smith and Schoene, 2012) in these circumstances, the use of computer gaming can give us a lot of hints and clues about the physical or mental impairments, and we can try to fix them by various exercises. These can also reduce their feelings of loneliness.

With the increase in age, there is a rapid increase of cognitive impairment. Alzheimer's disease is expanding day by. There are some evidences that cognitive training could improve cognitive function, which potentially slows cognitive decline and prevent the age-related cognitive problem such as dementia the earlier a cognitive impairment can be diagnosed, it is beneficial and can be treated by the products offered in this approach. In order to support active aging, various systems have been developed which can take out a solution to these problems:

1. CoCoMo:

It is a PC based cognitive ability measurement tool. It can measure cognitive abilities by testing the daily living problems such as money counting and etc. it is a very reliable method, and very easy to use. Anybody can get familiar with it easily.

2. E-Core:

   As a cognitive rehabilitation training system based on embodied cognition theory, E-CoRe System makes simultaneous training of cognition and physical movement possible.

Such systems are very beneficial for health issues and are very easy to use too. They don't provide the hindrance of unfamiliarity with a problem, rather they fix the problems in a healthy and enjoyable environment. These systems represent a new interactive therapeutic approach towards better cognitive rehabilitation.

The object of automatic speech recognition is to capture an acoustic signal representative of speech and determine the words that were spoken by pattern matching. Speech recognizers typically have a set of stored acoustic and language models represented as patterns in a computer database. The words spoken are then matched with the patterns already saved and hence it is very easy for the elderly people to manage many things or set reminders and do many other jobs with it.

However the problem with this device is that the speech patterns of people vary with age. Sometimes there are such waveforms, which are not audible by such devices. Hence it becomes difficult for them. A lot has to be worked on like the font size and graphic interface so that there are no such problems in the promotion of familiarity.

### 2.2.5 Use of games as a 'personalization' and 'familiarity approach

Gerontechnology can be defined as an interdisciplinary field of study of aging and technology so as to ensure good health, completeness in terms of social participation, and autonomous living through the entire lifetime of an individual. The reason why Gerontechnology is considered interdisciplinary in nature is that it combines gerontology characterized by medical, psychological, and social sciences of aging and technology comprising of robotics, ergonomics, and information and communication technologies. From this context, the elderly population is evaluated from the perspective of residing in a society that is technology driven, whereas technology is evaluated and investigated with a motive to be utilized as a means of improving the overall living standards of the elderly population, and increment their social engagement levels. Since the ageing / elderly population is an inevitable part of the society today, it has been contended that the quality and autonomy in living for this population is attainable by virtue of continuous technological innovations, which in turn would also cater business and economic development (Wu, et al, 2015).

From the perspective of technology use in the elder population, 2 main categories can be identified. The first category relates to the latest and innovative Information and Communication Technologies, whose target market is fairly broad. The second category encompasses the utilization of Assistive Technologies (ATs) specifically aimed at facilitating those experiencing or suffering from certain impairments or disabilities such as visual impairments, hearing loss, dementia / memory loss, difficulties in mobility and other such disabilities. Both of these categories, although different in terms of their intended target audience, have been put in place to serve the same purpose; improving the quality of life for individuals. Since there are several conditions apart from the cognitive decline and possible dementia, such as sensorial and physical declines and other health issues. In response to this potential demand, ICT-related products, such as robotics, smart home technology, assistive communication devices, and sensors for social alarms, are either under research and development or already on the market. These products are commonly referred to as gerontechnologies. Nonetheless, staying in context of the current thesis, gerontechnologies seem to be quite useful in developing and attaining personalization and familiarity among the elder population representatives experiencing cognitive decline or dementia. It is a common observation that the elder population feels comfortable in using the older technologies, but are very reluctant in adopting the new technologies as they lack engagement and knowledge. Under such circumstances, utilization of gerontechnology and interactive gamification can prove to be quite fruitful. Research has also proven that compared to the younger population, the elderly population has a lesser degree of access to the newer technologies (Olson et al, 2011). This reduced exposure, and the reluctance of older adults to adopt to newer technologies eventually leads to unfamiliarity with the new products, and may limit their cognitive as well as physical activities to lead an independent lifestyle. Provided the fact that new technologies are developed to support and make life easier than before, familiarity with these technologies is important

## 2.2.6 The potential of familiarity in game design

Video games play a role in the emotional and mental development of people. Games are played by any group of age, they are even played by the elderly people these days due to the reasons they need different kinds of activities in order to engage themselves in the old age houses as well as the nursing homes. By using game aesthetics and thinking group-based mechanics, gamification can engage people more strongly. (Kapp, 2012) co playing games can improve many other aspects of personality such as intrinsic motivation, cognitive apprenticeship and flow. (Dede, 2009)

People at the age of sixty are mostly at the verge of retirement or have already got retired. Ageism is a subtle demon. You have got nothing else to do than time to think about all the tensions and worries of the world. They all need something to keep busy in so that they can add some meaning to their dull and monotonous day. At this age one is usually under a lot of stress and some even suffer from psychological disorders because they worry too much. (Nap, Kort, IJsselsteijn,Poels, n.d.)

. Digital games and technology indeed, do have the skill of bringing joy, colors and fun to the dull lives. Psychological researches indicate that playing games have got usually a lot of benefits. Voices of nature can be soothing. (Despain, 2012) Digital games can provide them this opportunity and enjoyment by integrating some playful elements. For this purpose, different types of games have been designed varying according to the age, gender, interests and education level of people. But needs to be taken care of is that the games they play are environment friendly for the respective people rather than being absurd. (Crawford, 1997)

It is evident that the social, environmental and demographical trends tend towards the elderly, having a massive effect on the way the games should be designed for them. This makes it obvious for the game designers to take care of the minute details and

characteristics, which may help in bringing familiarity with the game for the elders. (Crawford, 1997)

The first and the foremost problem that the elderly people who play games encounter is that they have hearing problems. (Adams, 2013) Almost everywhere across the world in the elderly community, there is the problem of hearing senses. So the game designers should take great care of this issue right from now on because hearing- aids are going to become a very chief problem for them in the near future. () majority of the hearing aids that the elderly people use are due to the typical age-weakening hearing problems. So they can play a crucial role in the development of games. (Holzinger, Ziefel, Hits and Debevec, 2013). This can be avoided by adding audio devices to the games and they are being able to play it in many devices in different places.

There have also been introduced various other features such a team play, and avoiding the usage of unnecessary colors or creating a childish environment appearance. The games are also player centered and they offer learning on a very good level. Also, the reason we play games is in order to escape from our hectic daily routines or some sad realities. These games can also offer relax and tension free environment to the elderly people where they can play and forget their dull and sad routines. (Nap, Diaz-Orueta, Bierhoff, Heuvel, Mafreda, Dolnicar, 2012)

Usually the games are mainly designed for young adults because they are very much among the young generation; but now even the old age people have started participating and for this reason certain games can be found. (Nap, Kort, Ijsselsteijn, 2009)

According to a research held recently by the BBC, people with ages varying from 51-65 tend to play games that are mostly puzzle games or problem solving. Senior specific lifetime experiences, world knowledge, age-related changes in perception, cognition, and motor control are likely to have an influence on specific gaming preferences, their motivations and needs. They don't like racing or shooting guns and other games of the sort due to the reason that they neither fast enough, nor do they like violence, shouting and noise. (Nap, Kort, Ijsselsteijn, 2009)

Games usually played by people with ages above 70 haven't been done a research on, that what kind of games do they like to play; nor having any games with their category been designed. Although the number of the senior gamers is increasing, yet very little is known of what kind of games would they like to play or tend to play. It is quite unclear what motivates them to engage in digital gaming, what type of games do they like to play, and what problems they have with game interfaces while they are playing, and what their perceptions and attitudes are towards digital gaming. (Nap, Kort, Ijsselsteijn, 2009)

According to the Adams (2013), a five-factor model of VandenBerghe can also be applied to game designing, which involves the following points:

1. There should be novelty in the games. People like unexpected variety and a lot of experience.
2. There should be desire for challenge. And perhaps more specifically effort and control.
3. People like co playing and party gaming. Those who prefer to avoid stimulation games can play alone too.
4. Quality of the game: the bringing of all the parts of a game into a coherent whole.
5. An element of danger or fright.

A research was held on what elderly people had to say about the prevalent games in the market. Following are few of their views about it:

They are very horrible, and there is already so much misery in the world. They don't provide the environment, which can take us all away from such sad facts. There shouldn't be any death of the members, and many said they were against violence. Some reported that they couldn't understand the game rules. A few were of the view that they didn't feel quite comfortable with the situations in the game like beating granddad, or games being too glamorous.

Nearly all adults showed very strong emotions to the types of games that are about hatred and violence. Certainly, there is quite a capital level of difference in the interests of the young generation and the elderly community. If it is observed generally, we can find out that the elderly people like more of being wise and they act like it too. Therefore, the games that are to be designed for them should be more logical or problem solving rather than being irritating and creating a mess in their minds. Similarly, the rules on the games should be quite easier and the interface should be much simpler rather than the advance and complicated XBOX, where they find difficulty in how get into the next level and they get stuck on a specific level for lifetime. As a consequence of both functional changes and a lack of technological expertise, people of elder community are hurt more by usability problems than younger users. (Zugrich et al, 2013)

Older adults find it difficult to solve complex state of games, which make them divert their attention and lose their interest very soon. They have enough impairments and weaknesses of their own at such an age to attend to, rather than facing such higher-level problems. They may have chronic disorders, arthritis, heartaches and other diseases of the sort. So games should be designed for the sake of games only. And that is to offer them escapism from their worries. (Zugich et al, 2013)

The use of large and comprehendible symbols should be recommended. Also the auditory devices and problems should be taken care of. Flores et al. highlight the creation of meaningful play through learning objectives and social play. The Elder Games project results recommend that senior players have a preference for simple puzzles and quiz games, which may have a positive impact on daily life. The games should be designed to sharpen one's mind and that they should seek information from it. (Gerling, Schulte, Smeddinck and Masuch, 2012) and they should not only focus on the rules but there should be an additional set of objectives, which focuses on the personal growth and the interaction among people. This will make them feel highly good of themselves and others.

Generally, the availability of in-game resources should not be as restrictive as for a younger audience and ideally be individually adjustable and flexible. There should be a set of rules and objectives displayed in the beginning so that they are communicative enough for user and they can start the game without any such fear. (Gerling, Schulte, Smeddinck and Masuch, 2012)

In order to create a new game, the design of the game is of utmost importance and relevance. There should be additional focus groups and co designs so that there is no repetition, nor is there any kind of hindrance in the middle. (Gerling, Schulte, Smeddinck and Masuch, 2012). The stronger the design and in accordance with the age factors, the more entertaining it will be.

# Chapter 3 – The Effects of Familiarity in Design on the Adoption of Wellness Games by the Elderly

Research was carried out to investigate the impact of familiarity in design on the adoption of wellness games for the elderly. This research was published in the proceedings of the 2015 IEEE/WIC/ACM International Joint Conference on Web Intelligence and Intelligent Agent Technology (WI-IAT'15), 2015. We proposed a familiarity design framework with three dimensions of familiarity design: 1) *symbolic* familiarity; 2) *cultural* familiarity; and 3) *actionable* familiarity. We then conduct a focused group study involving 10 people over 65 years old to experience two wellness games containing different sets of with familiarity design elements. The results show that familiarity in design improves the perceived satisfaction and adoption likelihood significantly among the elderly users. With these results, we discuss how the study can benefit intelligent interface agent design when these agents need to interact with elderly users. (Yu et al., 2010 & Yu et al., 2007).

## 3.1 The framework for familiarity design

The effect of familiarity on the interaction between a user and a technological product can be summarized in three stages. The first stage relates to the works of Fitts and Posner [30], Taatgen et al [31], and Ericsson [32], and can be referred to as the "Cognitive stage". It revolves around the basic knowledge required for task execution. The second stage is referred to as the "Associative stage" in which the persons move from their declared understanding of the action to a process based performance of the action. The stronger the association with something the person is already familiar with, the higher the speed of execution and the lower the amount of effort needed to carry out the action [31]. The third stage is called the "Autonomous stage" in which the actions follow a smooth procedure and actions are quick and involve negligible effort.

Based on this three-stage model of the effect of familiarity, we propose a framework of familiarity design for human- computer interactions. It consists of three dimensions of familiarity design elements:

1) *Symbolic Familiarity*: objects, activities or processes commonly occurring in the target users' daily life are infused into the design of a system.

2) Cultural *Familiarity*: concepts, artifacts, patterns, traditions, or rituals commonly appearing in the target users' cultures are infused into the design of a system.

3) *Actionable Familiarity*: the acts of interacting with the symbolic familiarity elements and cultural familiarity elements in a system are similar to the acts of interacting with these elements in real life.

Depending on the degree of fidelity in replicating the symbolic and cultural familiarity objects in a system, and the similarity between interacting with these elements in the system com- pared to interacting with them in real life, a given computing system can be quantitatively evaluated based on these three dimensions of familiarity design.

## 3.2 The focus group study

The preliminary study involved 10 senior citizens aged 65 and above, in order to research on how their adoption of interactive digital media technologies is affected by design elements familiar to their daily life.

### 3.2.1 Study Design

The familiarity research was carried out at Joint NTU-UBC Research Centre of

Excellence in Active Living for Elderly (LILY). For the purpose of carrying out research on familiarity, two similar interactive game systems are selected. They are the Personal wellness and rehabilitation suite (PWRS) (Figures 3 and 4) and King Ping Pong (Figure 5).

Both games are virtual table tennis games. As playing table tennis is a familiar activity in the participants' daily life, both games are considered to contain symbolic familiarity. PWRS contains culturally familiar elements (e.g., the dragons in the background) to the study participants who are Singaporeans, while King Ping Pong does not contain such elements. Thus, PWRS is considered to contain cultural familiarity while King Ping Pong is not. In PWRS, players can use body gestures similar to playing table tennis to control the virtual paddle, whereas King Ping Pong requires the player to use a mouse to control the paddle. Therefore, PWRS is considered to contain actionable familiarity while King Ping Pong is not.

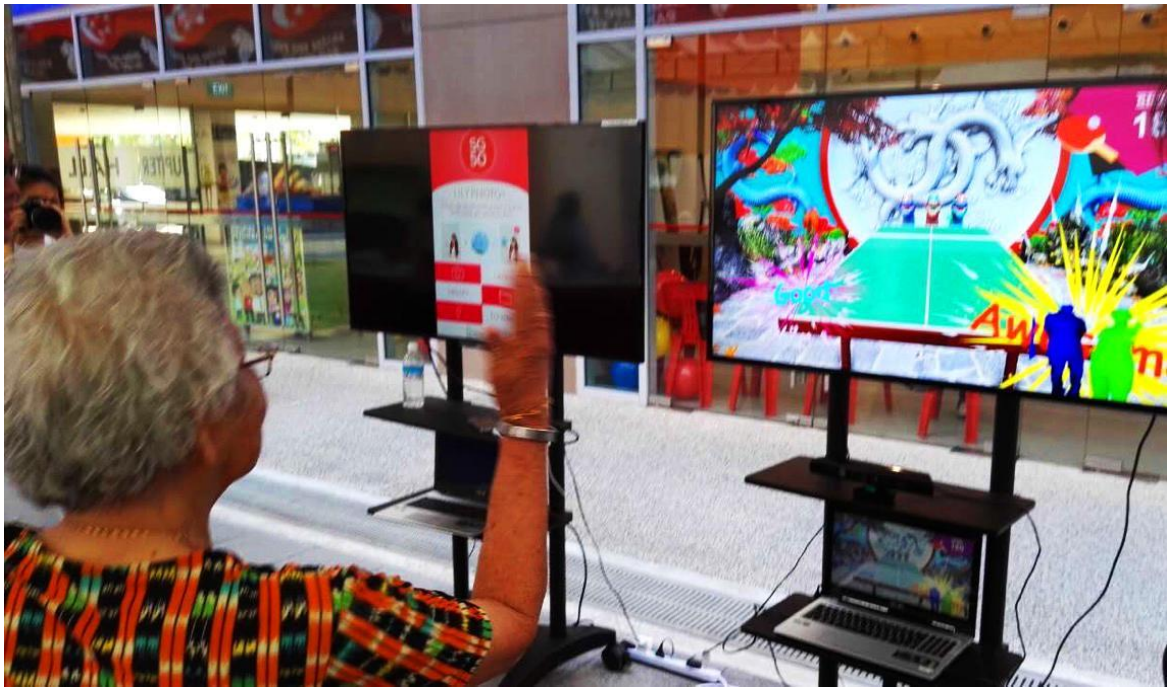

Figure 3: Observation of elderly's table tennis play involving camera with depth sensors (virtual reality)

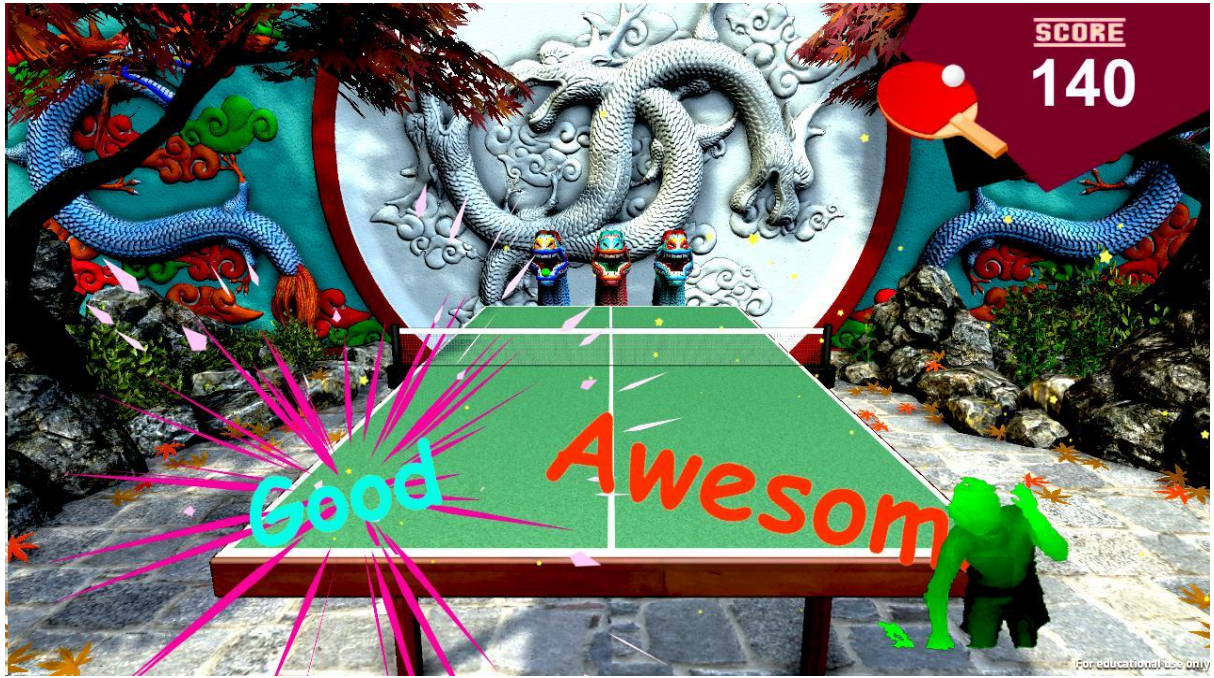

Figure 4: The PWRS table tennis game screen

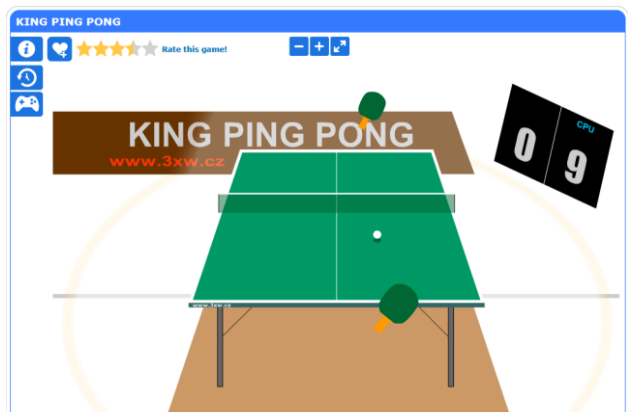

Figure 5: The King Ping Pong Game Screen

### 3.2.2 Study Participants

The study was carried out with the help of a focus group of 10 senior citizens with ages between 65 and 85 years with a mean age of 74.6 and a standard deviation of 6.80. They were asked to play both the PWRS game and King Ping Pong game. Each participant played both games uninterrupted for about 2 minutes each. They were observed during their game play sessions and interviewed at the end of each session. A survey form was filled up by each participant as part of the evaluation process. During this process, several aspects of gameplay were assessed. These included an assessment of the elderly's familiarity with contemporary games based on virtual reality such as table tennis, basketball and flying eagle. Their gameplay, emotional behavior, response to different situations during gameplay and their thoughts and comments were also observed and taken note of, for data analysis.

### 3.3 Results and analysis

In this section, we discuss the results obtained from analyzing the study participants' responses to the interviews and the survey questionnaire.

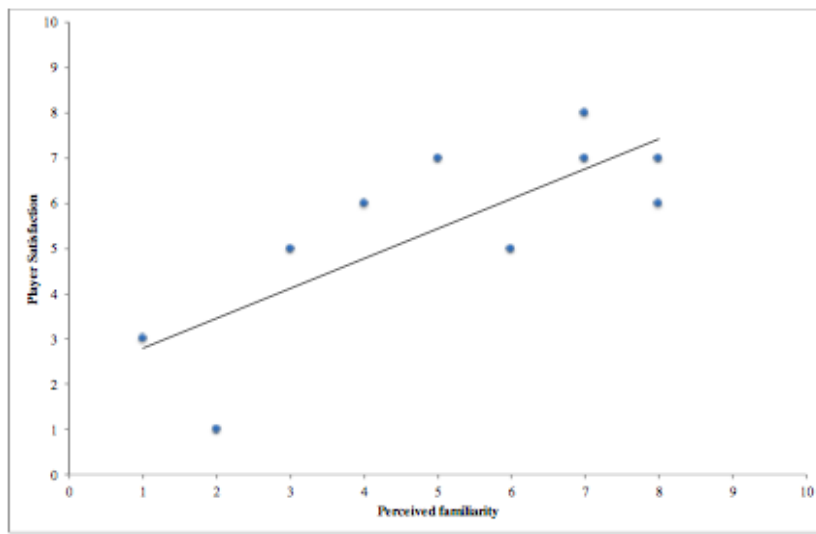

**Figure 6: Analysis of the relationship between satisfaction and perceived familiarity for PWRS**

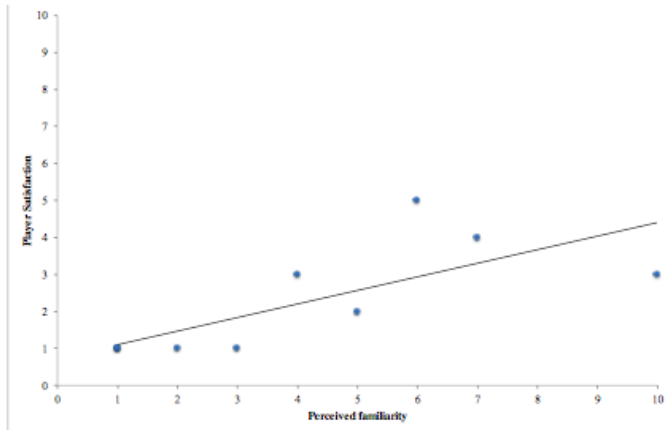

**Figure 7: Analysis of the relationship between satisfaction and perceived familiarity for King Ping Pong**

As can be seen from Figure 6, there is a strong positive correlation between the elderly participants' satisfaction and the perceived familiarity for PWRS (with a product moment correlation coefficient value of 0.78). Hence, it can be seen that the higher the perceived familiarity of a game, the greater the elderly user satisfaction is with the game.

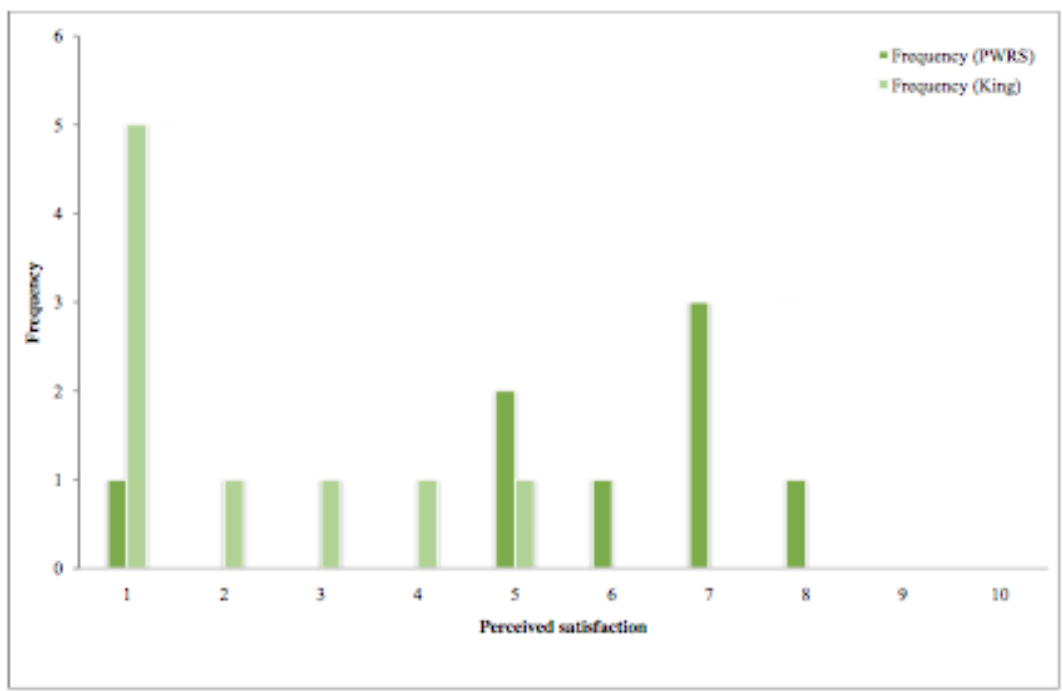

**Figure 8: Analysis for PWRS and King Ping Pong satisfaction**

As can be seen from Figure 7, there is a strong positive correlation between the elderly participants' satisfaction and the perceived familiarity for King Ping Pong (with a product moment correlation coefficient value of 0.75). Hence, the results obtained from Figure 6 are upheld. By Comparing Figure 6 with Figure 7, it is clear that player satisfaction for PWRS, which contains two more dimensions of familiarity design than King Ping Pong, is significantly higher than that for King Ping Pong.

It can be seen from the analysis of PWRS and King Ping

Pong games (Figure 8), the PWRS game is more popular among the participants and that the participants are more satisfied with it. The reason could again be linked to what is shown in Figures 6 and 7, that a greater perceived familiarity leads to a greater satisfaction with use.

From Figure 9, it can be seen from the analysis that there is a strong positive correlation between the likelihood of the adoption of a wellness game and the perceived familiarity with the wellness game (with a product moment correlation coefficient value of 0.82).

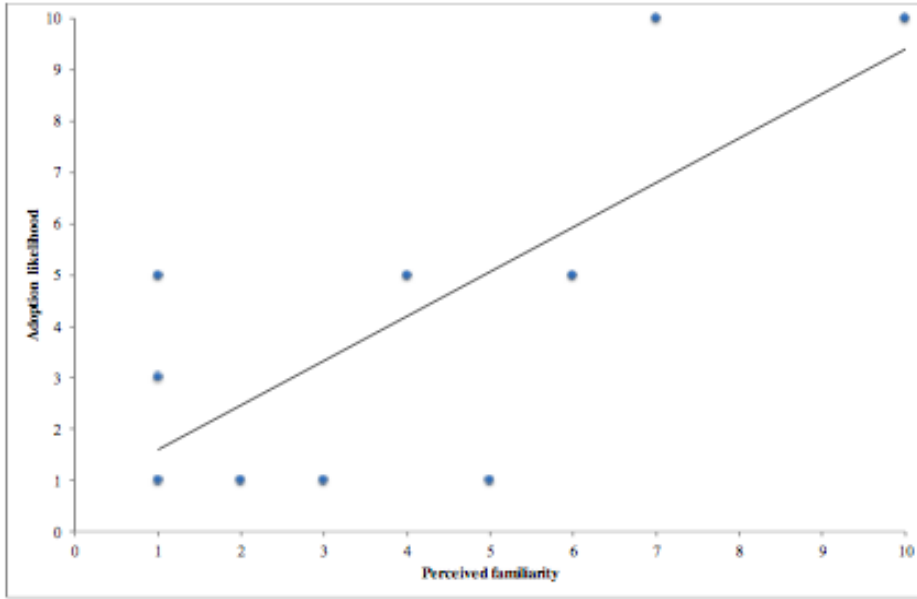

Figure 9: Analysis of relationship between adoption likelihood and perceived familiarity for King Ping Pong

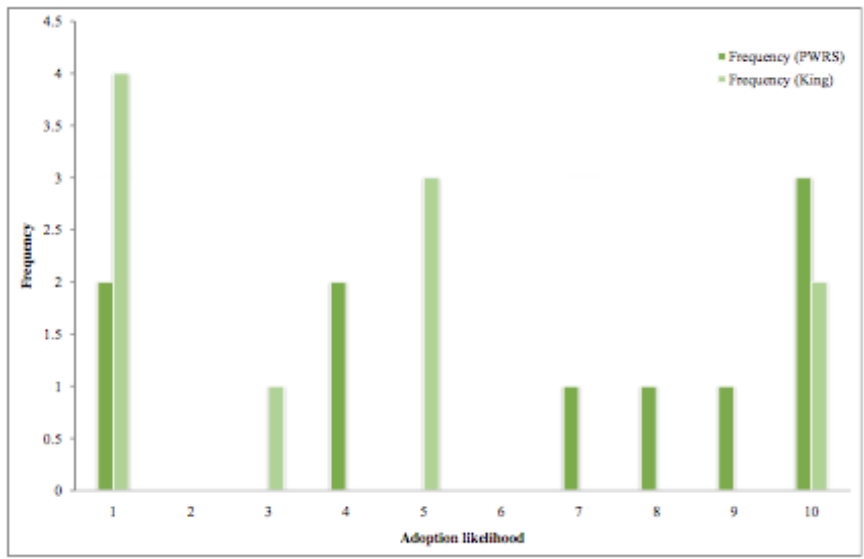

Figure 10: Analysis for adoption likelihood for PWRS and King Ping Pong

It can be seen from the analysis of King Ping Pong as well (Figure 10), that there is a strong positive correlation between the likelihood of the adoption of wellness games and the perceived familiarity with the wellness game (product moment correlation coefficient value of 0.74). Hence the greater the perceived familiarity, higher shall be the likelihood of adoption of the wellness game.

It can be seen from Figure 10, the PWRS game is more popular and that the participants are more likely to adopt it than King Ping Pong. The reason can again be linked to what is shown in Figures 9 and 10, that a greater perceived familiarity leads to a greater likelihood of adoption of the game.

## 3.4 Discussions and future work

This research has yielded novel results that a greater perceived familiarity with respect to a wellness game is likely to lead to a better performance in terms of the ability to do well in the game, a greater satisfaction with its use, and a higher likelihood of adoption by elderly users. This has important implications for the adoption of other technological platforms in the context of gerontechnology. With the proposed framework, researchers who need to design intelligent interface agents which interact with the elderly can have a method to analyze which dimension of familiarity design the agents require.

In future research, we will focus on two main areas. Firstly, we will conduct larger scale studies involving senior citizens from more diverse backgrounds to improve the generalizability of our results. We plan to achieve this objective through a crowdsourcing (Yu et al., 2012) based experimentation platform. Secondly, we will study the relative importance of the three dimensions of familiarity design to provide more detailed guidelines for researchers and developers serving the elderly.

# Chapter 4: Familiarity in productive aging

## 4.1 Towards efficient collaborative crowdsourcing

As was discussed in the literature review, it was ascertained that inter generational crowd sourcing had enormous potential for contributing towards addressing the needs of the elderly and making them a useful and important part of task completion. Hence, this research focused on finding an optimal way to bring the elderly and the young together to perform tasks that the elderly or the young cannot solve by themselves.

The elderly are retired and can't find jobs. However, the crowd-sourcing algorithm brings the elderly together and breaks bigger tasks into smaller ones. Then, through crowd sourcing, the elderly are employed in completion of tasks and they get paid, so the elderly are able to be productive and useful. Moreover, bigger tasks, that are challenging, are broken down into smaller tasks and the crowd sourcing algorithm utilizes the elderly and gives them those tasks to do that they are good at (because of skill or experience), hence solving the bigger task, while at the same time, making the elderly useful and productive. Moreover, there is inter generational crowd sourcing. For example, the elderly know dialects while the youth do not. So the elderly can perform those tasks that require the use of dialects and the youth can perform those tasks that need energy or being technologically savvy. By intergenerational collaboration through intergenerational crowd sourcing, bigger and complicated tasks are completed by the collaboration of the elderly and the youth, facilitated by technology and this algorithm - The crowd-sourcing algorithm.

We consider the problem of task allocation in collaborative crowdsourcing (e.g., inter-generational crowd- sourcing) systems in which different skills from multiple workers need to be combined to complete a task. We propose *CrowdAsm* - an approach which helps collaborative crowdsourcing systems determine how to com-bine the expertise of available workers to maximize the expected quality of results while minimizing the expected delays. Analysis proves that the proposed approach can achieve close to optimal profit for a crowdsourcing system if workers follow the recommendations.

In recent years, intelligent task allocation has been recognized as a useful approach to make efficient quality-time- cost trade-offs in crowdsourcing. Existing task allocation approaches in crowdsourcing focus on distributing tasks that can be performed by an individual worker (Yu et al. 2015). This is also the case for crowdsourcing with complex workflows (Tran-Thanh et al. 2015). As populations age, inter-generational crowdsourcing platforms start to emerge. Workers with diverse skills need to form effective teams to collaboratively complete crowdsourcing tasks requiring heterogeneous skills. An example is to proofread traditional Japanese scripts where elderly workers contribute linguistic expertise while the young assist them with technical knowhow (Kobayashi et al. 2013).

To address this problem, we propose the crowd assemble (CrowdAsm) approach1. It dynamically assembles teams of workers considering the budgets, the availability of workers with the required skills and their track records to complete crowdsourcing tasks requiring collaboration among workers with heterogeneous skills. It maximizes the expected quality of the task results, minimizes the expected time elapse, and stays within the given budget (Figure 1). Through rigorous analysis, we prove that CrowdAsm can achieve near optimal profit for a given collaborative crowdsourcing system if workers follow the recommendations.

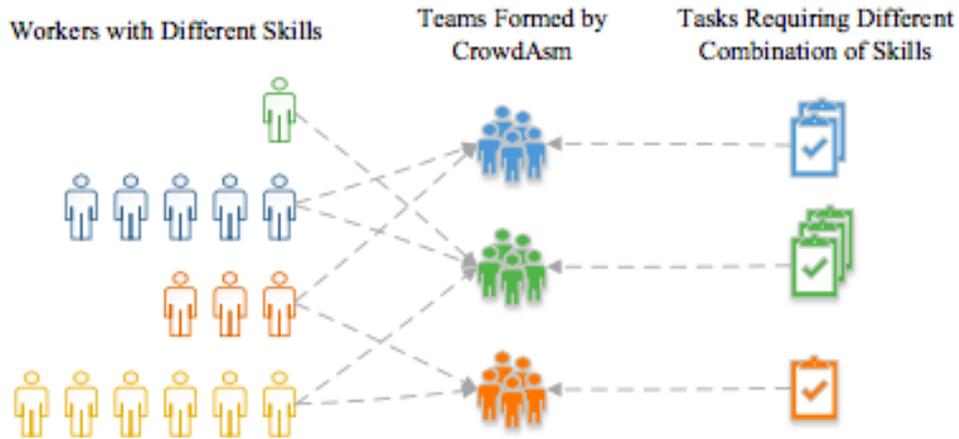

Figure 11: Crowdsourcing team formation for Crowd Asm

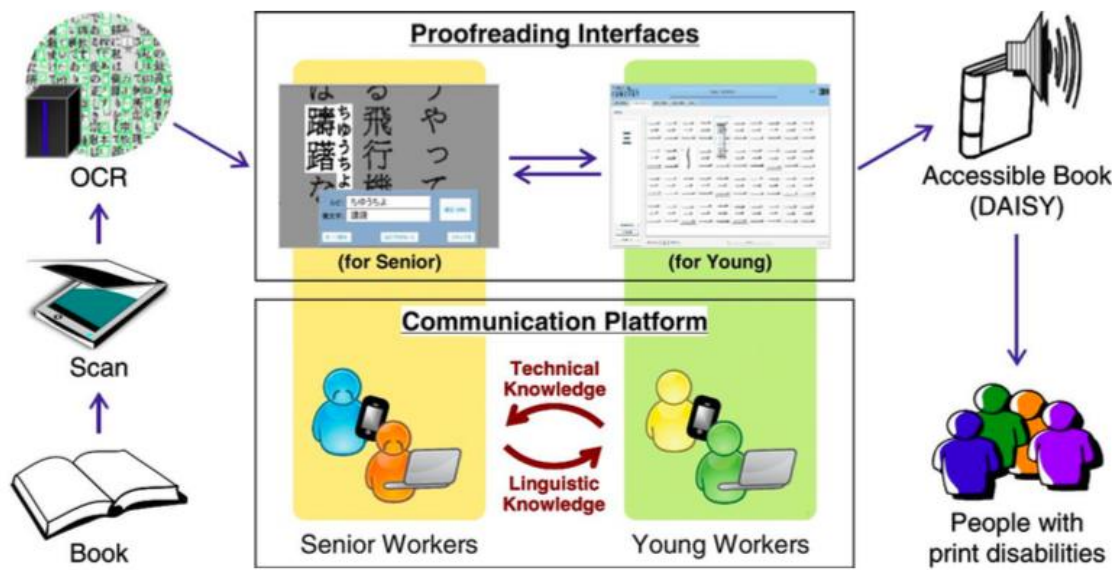

Figure 12: An overview of the inter generational collaborative crowdsourcing for proofreading of e-books in traditional Japanese scripts (Kobayashi et al., 2013)

### 4.1.2 Crowdsourcing system model

For a given crowdsourcing system, the availability of workers with a given skill m can be modeled as a queuing system:

$$q_m(t+1) = q_m(t) - \mu_m(t) + a_m(t) \tag{1}$$

$\mu_m(t) = \sum_{k=1}^{K} n_{m,k} \tilde{T}_k(t) - \tilde{a}_m(t)$ (Paper 3) where $n_{m,k}$ is the number of workers

with skill m required to complete a task of type k, and $\tilde{T}_k(t)$ is the actual number of tasks of type k completed during time step t. The range of values for $\tilde{T}_k(t)$ is $0 \leq \tilde{T}_k(t) \leq d_k(t) T_k(t)$. $T_k(t)$ is the number of tasks of type k demanded by the crowdsourcers at time step t. $d_k(t)$ is an indicator function reflecting the ability of the crowd-sourcing system to satisfy the demands for tasks of type k. It can be expressed as:

$$d_k(t) = \begin{cases} 1, & \text{if } q_m(t) \text{ can support } n_{m,k} > 0 \text{ for all } m \\ 0, & \text{otherwise.} \end{cases}$$

(2) Due to the physical limitations of a crowdsourcing system, we assume that there exists an upper bound to $\mu_m(t)$ which can be expressed $\mu_m^{max}(t) = \sum_{k=1}^{K} n_{m,k} T_k^{max}(t)$. $\tilde{a}_m(t)$ is the number of workers with skill m who have become available at time step t (e.g., those who have completed previous tasks allocated to them or new workers joining the system).

$a_m(t) \in Z^+ \cup \{0\}$, is the number of workers to be mobilized by the CrowdAsm approach. For a given crowdsourcing system, the number for registered workers is finite. Thus, we assume that there is an upper bound $a_m^{max}$ for $a_m(t)$ such that $0 \leq a_m(t) \leq \bar{a}_m$. The cost incurred for mobilizing a vector $\mathbf{a}(t)$ of workers under the supply condition $\mathbf{x}(t)$ is:

$$c(\mathbf{a}(t), \mathbf{x}(t)) \triangleq \sum_{m=1}^{M} \left[\frac{c_m(t) a_m(t)}{\tilde{r}_m(t)}\right] \quad (3)$$

where $c_m(t)$ is the cost required to attract workers with skill m to become available under a given worker supply condition $x_m(t)$. Apart from the financial cost, the quality of the available workers as measured by an average reliability value $0 < \tilde{r}_m(t) < 1$ is also taken into account. Such a measure $r_i(t)$ for each worker i can be computed based on his past performance using reputation-based approaches such as (Jøsang and Ismail 2002). Thus,

$$\tilde{r}_m(t) = \frac{1}{N_m^{out}(t)} \sum_{i=1}^{N_m^{out}(t)} r_i(t) \quad (4)$$

where $N_m^{out}(t)$ is the total number of workers with skill m who are not currently logged in to the crowdsourcing system and whose $r_i(t) \geq \varepsilon$. $\varepsilon$ is a reliability threshold specified by the crowdsourcing system or a crowdsourcer (e.g., the minimum "approval rate" in mTurk).

The demand $T_k(t)$ for tasks of type k from crowdsourcers may be affected by the workers' performance and the current price the crowdsourcing system charges for this type of tasks $p_k(t)$. The demand can be denoted by the function $f_k(p_k(t), \tilde{r}_k(t))$. Recent large scale empirical studies involving sellers from both eBay and Taobao (Ye et al. 2013) yielded an expression relating new workload for a seller's to his price and reputation. In this study, we use the result from the aforementioned empirical study to

model the demand for tasks of type k as:

$$\ln f_k(p_k(t), \tilde{r}_k(t)) = \alpha_1 + \alpha_2 \ln \tilde{r}_k(t) + \alpha_3 \ln n^{(+)}(t) + \ln p_k(t). \quad (5)$$

$\alpha_1$ to $\alpha_3$ are constant values. Based on the regression analysis results from (Ye et al. 2013), $\alpha_2 < 0$ and $\alpha_3 > 0$.

$n^{(+)}(t)$ denotes the number of positive ratings received by k

the crowdsourcing system for helping crowdsourcers complete tasks of type k over a given period of time. In sub- sequent parts of this paper, we adopt Eq. (5) to model the relationship between the demand for tasks of type k and the average reliability measure of the workers involved $\tilde{r}_k(t)$ as well as the $p_k(t)$ value. [1] Thus, $\mu_m(t)$ can be re-expressed as

$$\mu_m(t) = \sum_{k=1}^{K} n_{m,k} f_k(p_k(t), \tilde{r}_k(t)) - \tilde{a}_m(t) \quad (6)$$

### 4.1.3 Derivation of CrowdAsm
The expected profit for the crowdsourcing system at time step t can be expressed as:

$$\delta(t) = \sum_{k=1}^{K} d_k(t) f_k(p_k(t), \tilde{r}_k(t)) p_k(t) - c(\mathbf{a}(t), \mathbf{x}(t)). \quad (7)$$

Using the *Lyapunov drift* $\Delta(q(t))$ (Neely 2010) as a metric to the overall level of congestion of demand on workers with various skills in the crowdsourcing system (i.e., expected de- lay), we formulate the *delay-minus-profit* objective function as:

$$\Delta(\mathbf{q}(t)) - \rho \mathbb{E}\{\delta(t)|\mathbf{q}(t)\} \quad (8)$$

where ρ > 0 is a weight factor determining the relative importance of delay and profit.

According to the definition of Lyapunov drift:

$$\Delta(\mathbf{q}(t)) \triangleq L(\mathbf{q}(t+1)) - L(\mathbf{q}(t))$$

$$= \frac{1}{2}[\sum_{m=1}^{M} q_m^2(t+1) - \sum_{m=1}^{M} q_m^2(t)]$$

$$= \frac{1}{2}\sum_{m=1}^{M}(a_m(t) - \mu_m(t))^2 + \sum_{m=1}^{M} q_m(t)(a_m(t) - \mu_m(t))$$

$$= \frac{1}{2}[a_m^2(t) + \mu_m^2(t)] - a_m(t)\mu_m(t)$$

$$+ \sum_{m=1}^{M} q_m(t)(a_m(t) - \mu_m(t))$$

$$\leq \frac{1}{2}[(a_m^{max}(t))^2 + (\mu_m^{max}(t))^2] - a_m(t)\mu_m(t)$$

$$+ \sum_{m=1}^{M} q_m(t)(a_m(t) - \mu_m(t))$$

$$\leq \frac{1}{2}[(a_m^{max}(t))^2 + (\mu_m^{max}(t))^2]$$

$$+ \sum_{m=1}^{M} q_m(t)(a_m(t) - \mu_m(t))$$

(9)

where $L(\cdot)$ is the *Lyapunov function*. Let $\xi = \frac{1}{2}\sum_{m=1}^{M}[(a_m^{max}(t))^2 + (\mu_m^{max}(t))^2]$, we have:

$$\Delta(\mathbf{q}(t)) \leq \xi + \sum_{m=1}^{M} q_m(t)(a_m(t) - \mu_m(t)) \quad (10)$$

By taking exponent on both sides of Eq. (5), we have:

$$f_k(p_k(t), \tilde{r}_k(t)) = e^{\alpha_1 + \alpha_2 \ln \tilde{r}_k(t) + \alpha_3 \ln n_k^{(+)}(t) + \ln p_k(t)}$$

$$= \alpha_2 \alpha_3 n_k^{(+)}(t) e^{\alpha_1} \tilde{r}_k(t) p_k(t)$$

(11)

Let $\beta_k = \alpha_2 \alpha_3 n_k^{(+)}(t) e^{\alpha_1}$, Eq. (11) becomes:

$$f_k(p_k(t), \tilde{r}_k(t)) = \beta_k \tilde{r}_k(t) p_k(t) \quad (12)$$

Thus, Eq. (8) can be re-expressed as:

$$\Delta(\mathbf{q}(t)) - \rho \mathbb{E}\{\delta(t)|\mathbf{q}(t)\} \leq$$

$$\xi + \sum_{m=1}^{M} q_m(t)(a_m(t) - \mu_m(t))$$

$$+ \rho \sum_{k=1}^{K} \mathbb{E}\{d_k(t) f_k(p_k(t), \tilde{r}_k(t)) p_k(t)|\mathbf{q}(t)\}$$

$$- \mathbb{E}\{c(\mathbf{a}(t), \mathbf{x}(t))|\mathbf{q}(t)\}$$

$$= \xi + \sum_{m=1}^{M} q_m(t)[\mathbb{E}\{a_m(t)|\mathbf{q}(t)\}$$

$$- \sum_{k=1}^{K} n_{m,k} \mathbb{E}\{f_k(p_k(t), \tilde{r}_k(t))|\mathbf{q}(t)\} + \mathbb{E}\{\tilde{a}_m(t)|\mathbf{q}(t)\}]$$

$$+ \rho \sum_{k=1}^{K} \mathbb{E}\{d_k(t) f_k(p_k(t), \tilde{r}_k(t)) p_k(t)|\mathbf{q}(t)\}$$

$$- \rho \mathbb{E}\{c(\mathbf{a}(t), \mathbf{x}(t))|\mathbf{q}(t)\}$$

$$= \xi + \sum_{m=1}^{M} q_m(t)[\mathbb{E}\{(a_m(t)|\mathbf{q}(t)\}$$

$$- \sum_{k=1}^{K} n_{m,k} \mathbb{E}\{\beta_k \tilde{r}_k(t) p_k(t)|\mathbf{q}(t)\} + \mathbb{E}\{\tilde{a}_m(t)|\mathbf{q}(t)\}]$$

$$+ \rho \sum_{k=1}^{K} \mathbb{E}\{d_k(t)[\beta_k \tilde{r}_k(t) p_k(t)] p_k(t)|\mathbf{q}(t)\}$$

$$- \rho \sum_{m=1}^{M} \mathbb{E}\{\frac{c_m(t) a_m(t)}{\tilde{r}_m(t)}|\mathbf{q}(t)\}$$

(13)

By choosing only terms containing $a_m(t)$ from Eq. (13), we have:

Minimize:

$$\sum_{m=1}^{M} a_m(t)[q_m(t) - \rho \frac{c_m(t)}{\tilde{r}_m(t)}] \qquad (14)$$

Subject to:

$$a_m(t) \leq a_m^{max} \qquad (15)$$
$$r_i(t) \geq \varepsilon, \forall i \in a_m(t) \qquad (16)$$
$$\sum_{m=1}^{M} c_m(t) a_m(t) \leq B(t) \qquad (17)$$

$B(t)$ is the budget available for a given task.

To minimize Eq. (14), compute the values of the expression $[q_m(t) - \rho \frac{c_m(t)}{\tilde{r}_m(t)}]$ for all $m$. Sort all task requests in ascending order of their $[q_m(t) - \rho \frac{c_m(t)}{\tilde{r}_m(t)}]$ values. For each group of workers with skill $m$, send task requests to as many workers $a_m(t)$ as allowed by Constraints (15) to (17).

### 4.1.4 The CrowdAsm Algorithm

Based on the mathematical derivation in the previous sections, the CrowdAsm algorithm is as listed in Algorithm 1.

---

**Algorithm 1** CrowdAsm

**Require:** $q_m(t)$; $r_i(t)$ for each worker $i$; $\rho$, the set of task requests $\mathbf{T}_k(t)$; $B(t)$; $N_m^{in}(t)$, the total number of workers with skill $m$ who satisfy $r_i(t) \geq \varepsilon$ and currently logged in for all $m$; and $c_m(t)$ for all $m$.

1: Compute all $\tilde{r}_m(t)$ values according to Eq. (4).
2: Arrange task requests $k \in \mathbf{T}_k$ in ascending order of their $[q_m(t) - \rho \frac{c_m(t)}{\tilde{r}_m(t)}]$ values.
3: **for** (each worker skill $m$ required to complete tasks of type $k$) **do**
4:   **if** $(N_m^{in}(t) < n_{m,k} T_k(t))$ **then**
5:     $a_m(t) = 0$.
6:     **for** (each worker $i$ with skill $m$ who is not currently logged in **and** $r_i(t) > \varepsilon$ **and** $a_m(t) < n_{m,k} T_k(t) - N_m^{in}(t)$ **and** $B(t) \geq c_m(t)$) **do**
7:       $a_m(t) \leftarrow a_m(t) + 1$.
8:       $B(t) \leftarrow B(t) - c_m(t)$.
9:       Add $i$ into the set of candidate workers with skill $m$, $\mathbf{W}_m(t)$.
10:     **end for**
11:     **if** $(a_m(t) + N_m^{in}(t) \geq n_{m,k} T_k(t))$ **then**
12:       **for** (all $i \in \mathbf{W}_m(t)$) **do**
13:         Send out the task request for $i$ via the preferred channel of communication.
14:       **end for**
15:     **else**
16:       $d_k(t) = 0$.
17:     **end if**
18:   **end if**
19: **end for**

---

### 4.1,5. Analysis

In this section, we analyze the performance of the CrowdAsm approach. Specifically,

we are interested in the proximity of the time averaged profit for the crowdsourcing system to the optimal time averaged profit.

CrowdAsm observes the queues of available workers with different skills q(t) at any time step t and helps the crowd- sourcing system determine the values of a(t) to minimize Eq. (13). There exist at least a combination of $a^*_m(t)$ and $\mu^*_m(t)$ values which satisfy all constraints and produce the optimal time averaged profit $E\{\delta^*(t)|q(t)\} = \delta^{opt}$ for the crowdsourcing system with a given worker population such that:

$$\Delta(\mathbf{q}(t)) - \rho\mathbb{E}\{\delta(t)|\mathbf{q}(t)\} \leq \xi - \rho\mathbb{E}\{\delta^*(t)|\mathbf{q}(t)\} \\ + \sum_{m=1}^{M} q_m(t)\mathbb{E}\{a^*_m(t) - \mu^*_m(t)|\mathbf{q}(t)\}. \quad (18)$$

As the optimal policy results in $\mathbb{E}\{a^*_m(t) - \mu^*_m(t)|\mathbf{q}(t)\} = 0$ for all $m$, we have:

$$\Delta(\mathbf{q}(t)) - \rho\mathbb{E}\{\delta(t)|\mathbf{q}(t)\} \leq \xi - \rho\delta^{opt}. \quad (19)$$

Following the definition of $\Delta(\mathbf{q}(t))$ and taking expectations on both sides of the above expression:

$$\mathbb{E}\{L(\mathbf{q}(t+1))\} - \mathbb{E}\{L(\mathbf{q}(t))\} - \rho\mathbb{E}\{\delta(t)\} \leq \xi - \rho\delta^{opt}. \quad (20)$$

Summing the above expression over all time steps $t \in \{0, 1, ..., \tau - 1\}$ yields:

$$\sum_{t=0}^{\tau-1}[\mathbb{E}\{L(\mathbf{q}(t+1))\} - \mathbb{E}\{L(\mathbf{q}(t))\}] - \rho \sum_{t=0}^{\tau-1}\mathbb{E}\{\delta(t)\}$$

$$= \mathbb{E}\{L(\mathbf{q}(\tau))\} - \mathbb{E}\{L(\mathbf{q}(0))\} - \rho \sum_{t=0}^{\tau-1}\mathbb{E}\{\delta(t)\}$$

$$\leq \tau\xi - \tau\rho\delta^{opt}. \qquad (21)$$

As $\rho > 0$, dividing both sides of the above expression by $\tau\rho$ results in:

$$\frac{1}{\tau}\sum_{t=0}^{\tau-1}\mathbb{E}\{\delta(t)\} \geq \delta^{opt} - \frac{\xi}{\rho} + \frac{1}{\tau}\mathbb{E}\{L(\mathbf{q}(\tau))\} - \frac{1}{\tau}\mathbb{E}\{L(\mathbf{q}(0))\}. \qquad (22)$$

Since $L(\mathbf{q}(t)) \geq 0$ and $L(\mathbf{q}(0)) = 0$, we have:

$$\frac{1}{\tau}\sum_{t=0}^{\tau-1}\mathbb{E}\{\delta(t)\} \geq \delta^{opt} - \frac{\xi}{\rho}. \qquad (23)$$

Therefore, we have proven that the time averaged profit achievable for a given crowdsourcing system following CrowdAsm is within O($1/\rho$) of the optimal profit, subject to the physical limitations of the crowdsourcing system.

### 4.1.6. Conclusion
Hence by using the CrowdAsm we are able to dynamically assemble teams of workers considering the budgets, the availability of workers with the required skills and their track records to complete crowdsourcing tasks requiring collaboration among workers with heterogeneous skills. Theoretical analysis has shown that CrowdAsm can achieve close to optimal profit for a collaborative crowdsourcing system if workers follow the recommendations.

### 4.2. Productive Aging through Intelligent Personalized Crowdsourcing
It is important to make efficient quality-time-cost trade-offs in large-scale crowdsourcing networks through intelligent personalized task delegation/sub-delegation. In recent years, algorithmic crowdsourcing techniques have been proposed to help crowdsourcers delegate/sub-delegate tasks among workers efficiently taking into account situational factors such as their reputation and current workload. However, existing approaches assume that workers will accept the prices determined by the crowdsourcers and are not able to adjust the prices they charge for their services in response to changes in the situational factors facing them.

For this research it was proposed a reputation-aware task sub-delegation approach with dynamic worker effort pricing (Crowd Asm) to relax this assumption. By jointly considering a worker's current reputation, workload, and his trust relationships with others, Crowd Asm helps him make sub-delegation and pricing decisions in a distributed manner. The resulting task allocation efficiently utilizes crowdsourcing workers' collective capacity, provides provable performance guarantees, and significantly improves the workers' income. A short video explaining the motivations and main ideas behind crowdsourcing for productive aging can be found at https://www.youtube.com/watch?v=Epu1tWpmq-Y.

Crowd Asm based personal assistant agent to support productive aging through intelligent personalized crowdsourcing. The app is the result of collaboration between the research team and a local senior care community volunteer organization in Singapore. The objective of the app is to elicit the help from senior citizens to share their personal experience and stories in relation to given historical events, places, personalities, and artifacts based on photos contributed by local residents. For the time being, a point-based system is used in the app to price workers' effort instead of involving real money.

Figure 1 shows example screenshots from the app demonstrating functionalities including: 1) understanding user characteristics based on their profile information, 2) recommending photos to a user automatically by the Crowd Asm agent, and 3) example actions a user can perform on the recommended photos. The Crowd Asm agent intelligently recommends new tasks to users based on their personal situations to make efficient quality-time-cost trade-offs while avoiding overloading any workers involved.

Figure 13: Screenshots from the app

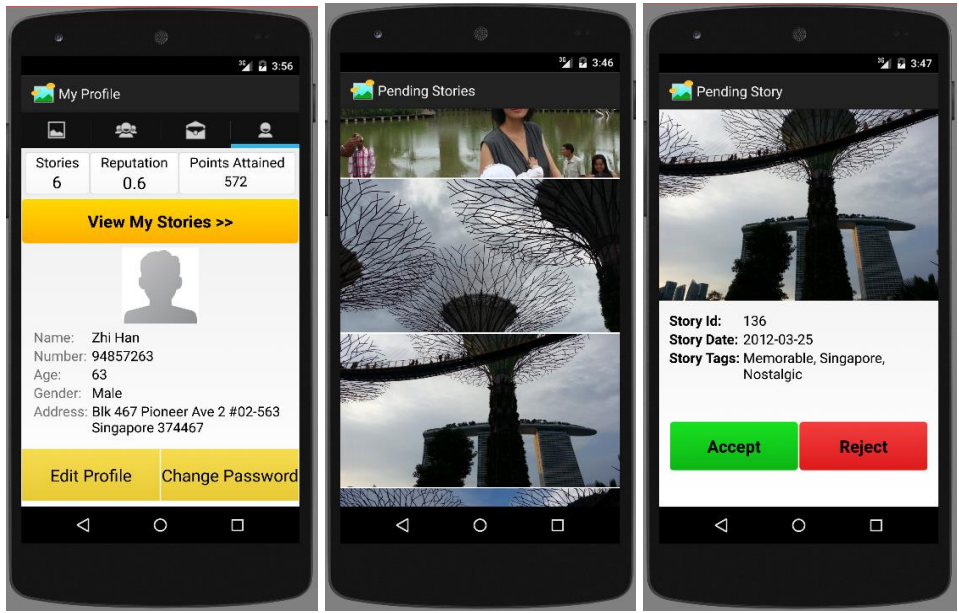

Figure 14: Recommendations made by Crowd Asm agents from multiple workers' apps as viewed from the server-side

Figure 2 visualizes a snapshot of the recommendations made by the Crowd Asm agents belonging to different workers at a given point in time. Statistical information related to each worker's characteristics and the situation he/she faces at that given instance (e.g., availability, reputation, workload, etc.) is shown. A link to an automatically generated short text explaining the rationale behind each recommendation can be accessed by clicking on the "View Explanations" hyperlink. The explanations are also available for the workers to view in order to facilitate understanding of the rationale behind the recommendations and help build trust between the workers and the Crowd Asm agents.

### 4.3 Towards Agent Augmented Inter-generational Crowdsourcing

#### 4.3.1 Demonstration content

In this research, we attempt to bridge this important gap with an inter-generational crowdsourcing based persuasive technique study platform - the SG50 Wish App. The platform consists of a mobile app and web-based platforms. It provides a channel for Singapore residents to share their well wishes in celebration of Singapore's 50th Anniversary of independence through multi-modal expressions including texts, photos, and augmented reality images.

The app is incorporated with a function that potentially requires inter-generational interactions. The augmented reality function of the app is able to detect an "SG50" logo through the camera embedded in a mobile device. Once the logo is detected, the app super-imposes a virtual birthday cake onto the logo to create a novel photograph for the user. However, this function requires two persons to cooperate. As young people are often more technically savvy than the elderly, they tend to cooperate in the way as illustrated in Figure 1.

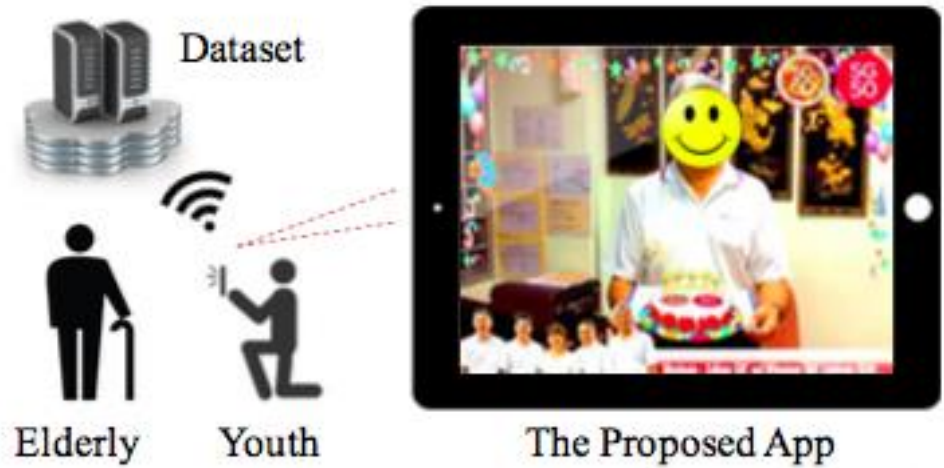

Figure 15: The proposed crowdsourcing-based app for investigating factors contributing to inter-generational interaction

We have embedded various central and peripheral persuasive designs into the mobile app based on the Elaboration Likelihood Model (ELM) persuasion theory (Cacioppo et al., 1986). These design factors include:

1. *The Appeal of National Spirit*: the app is positioned as a platform for people to share their wishes for celebrating the 50th National Day of Singapore;
2. *The Appeal of Sense of Belonging*: as a sub-function of the app, it allows people to write well-wishes for various participating community organizations;
3. *The Appeal of Social Networking*: the app and the web platform allows people to comment and like each others' photos and wishes;
4. *The Appeal of Innovative Interaction*: the app is incorporated with augmented reality functions such that it can produce specialized photos incorporated with various virtual artifacts.

The platform records sequence of usage for each user, the contents contributed by each users, and situational information such as time and location of usage and send these data into a central cloud server for storage. Together with the profile information (e.g., gender, age, cultural background) provided by the users during registration, we aim to build a large-scale dataset reflecting the relationship between people's background and their response to different persuasion techniques.

The system has been deployed in Singapore in collaboration with various local community organizations to start data col- lection. This dataset will allow us to investigate the motivating factors for people from diverse demographic backgrounds to engage in interactions with those from different age groups. The resulting dataset will be published to support the AI and MAS research community. This dataset will help researchers determine which of the persuasion mechanisms is likely to be in effect for each user, and act as an evaluation bench- mark for proposed agent-based solutions for motivating inter- generational interactions in crowdsourcing.

In subsequent research, we aim to propose models capturing the possible temporal dynamics inter-generational interaction behaviours. We will also leverage the goal oriented modeling methods (Yu et al., 2007) to represent complex knowledge possessed by the elderly to support inter-generational crowdsourcing of the elderly's wisdom and experience.

**Chapter 5 – Future Plan**

**5.1 Developing a Gamified Platform for Studying the Relative Importance of Product Familiarity.**

**5.2 Developing a Gamified Platform for Studying the Relative Importance of Process Familiarity.**

**5.3 Enhance a Situation-aware Crowdsourcing Team Formation Algorithm to Support Inter-generational Crowdsourcing.**

Singapore memory project: we want the elderly to contribute photos about Singapore. These will contribute towards Singapore memory project.

**5.4 Apple Game (water wheel - wellness game)**

I have also spoken to the Apple ResearchKit Team and they have expressed a keen interest in how we apply familiarity research. This is a good platform for the millions of Apple users to contribute towards medical research including various aspects of familiarity and personalization.

**5.5 Future Directions for this Research**

I will remain open and continue to observe how familiarity can contribute towards research pertaining to the elderly.

## 6. Conclusion

This research has tackled the imperative issues of catering to the elderly's health needs and of utilizing their skills for productive contribution to the society, with the help of crowdsourcing and intergenerational crowdsourcing. The important of familiarity and crowdsourcing has been established right from the literature review. This research builds the current lack of empirical studies on how the concept of familiarity can be infused into the design of interactive technology systems to bridge the digital divide preventing today's elderly people from actively engaging with such technologies. In this paper, a multi pronged approach is utilized. We researched the effects of familiarity in design on the adoption of wellness games by the elderly, familiarity in productive ageing, efficient collaborative crowdsourcing, productive ageing through intelligent personalized crowdsourcing and agent augmented inter-generational crowdsourcing. The results show that familiarity in design improves the perceived satisfaction and adoption likelihood significantly among the elderly users. These results can potentially benefit intelligent interface agent design when such agents need to interact with elderly users. A Crowdsourcing algorithm, CrowdAsm is developed. By using CrowdAsm we are able to dynamically assemble teams of workers considering the budgets, the availability of workers with the required skills and their track records to complete crowdsourcing tasks requiring collaboration among workers with heterogeneous skills. Theoretical analysis has shown that CrowdAsm can achieve close to optimal profit for a collaborative crowdsourcing system if workers follow the recommendations. Hence it has been shown that research on familiarity, personalization and crowdsourcing is highly beneficial to addressing the health needs of the elderly. It is even more crucial in ensuring that the elderly remain active and continue to contribute to solving the challenges faced by society; challenges that the youth cannot accomplish by themselves. Hence it is crucial that this research be continued further in order to ensure that familiarity research and inter-generational crowdsourcing be explored further.

# References

Algarabel, S., Rodrıguez, L. A., Escudero, J., Fuentes, M., Peset, V., Pitarque, A., et al. (2010). Recognition by Familiarity Is Preserved In Parkinson's Without Dementia and Lewy-Body Disease. *Neuropsychology , 24* (5), 599–607.

Araujo, R. M. (2013, March). 99designs: An analysis of creative competition in crowdsourced design. In First AAAI conference on Human computation and crowdsourcing.

Arch, A. (2008). Web Accessibility for Older Users: A Literature Review. *W3C Working Draft* .

Barry, J. E. (2008). *Every day habits and routines:design strategies to individualize home modifications for older people.* Washigton state university, Department of Interior Design.

Blackler, A., Mahar, D., & Popovic, V. (2010, November). Older adults, interface experience and cognitive decline. In *Proceedings of the 22nd Conference of the Computer-Human Interaction Special Interest Group of Australia on Computer-Human Interaction* (pp. 172-175). ACM.

Boger, J., Craig, T., & Mihailidis, A. (2013). Examining the impact of familiarity on faucet usability for older adults with dementia. *BMC Geriatrics , 13* (63).

Bressler, S. L., & Ding, M. (2006). Event-related potentials. *Wiley Encyclopedia of Biomedical Engineering* .

Cacioppo, J. T., Petty, R. E., Kao, C. F., & Rodriguez, R. (1986). Central and peripheral routes to persuasion: An individual difference perspective. *Journal of personality and social psychology*, *51*(5), 1032.

Cai, Y., Shen, Z., Liu, S., Yu, H., Han, X., Ji, J., Miao, C. (2014). *An Agent-based Game for the Predictive Diagnosis of Parkinson's Disease.* Paper presented at the 13th International Conference on Autonomous Agents and Multi-Agent Systems (AAMAS'14).

Domazet, DS., MC Yan, CFY Calvin, HPH Kong, and A Goh, "An infrastructure for inter-organizational collaborative product development System Sciences," In *Proceedings of the 33rd Annual Hawaii International Conference on System Sciences*, 2000.

Difallah, D. E., Catasta, M., Demartini, G., & Cudré-Mauroux, P. (2014, May). Scaling-up the Crowd: Micro-Task Pricing Schemes for Worker Retention and Latency Improvement. In *Second AAAI Conference on Human Computation and Crowdsourcing*.

Chrysikou, E. G., Giovannetti, T., Wambach, D. M., Lyon, A. C., Grossman, M., & Libon, D. J. (2011). The importance of multiple assessments of object knowledge in semantic dementia: The case of the familiar objects task. *Neurocase* , *17* (1), 57–75.

Cugelman, B. (2013). Gamification: What It Is and Why It Matters to Digital Health Behavior Change Developers. *JMIR Serious games* , *1* (1), e3.

Demirbileka, O., & Demirkanb, H. (2004). Universal product design involving elderly users: a participatory design model. *Applied Ergonomics* , *35*, 361–370.

Doan, A., Ramakrishnan, R., & Halevy, A. Y. (2011). Crowdsourcing systems on the world-wide web. *Communications of the ACM*, *54*(4), 86-96.

Gao, X. A., Bachrach, Y., Key, P., & Graepel, T. (2012, June). Quality Expectation-Variance Tradeoffs in Crowdsourcing Contests. In *AAAI*.

Giovannetti, T., Sestito, N., Libon, D. J., Schmidt, K. S., Gallo, J. L., Gambinoa, M., et al. (2006). The influence of personal familiarity on object naming, knowledge, and use in dementia. *Archives of Clinical Neuropsychology*, *21*, 607–614.

Gulati, R., & Sytch, M. (2008). Does familiarity breed trust? Revisiting the antecedents of trust. *Managerial and Decision Economics*, *29* (2-3), 165-190.

Jacques, J. T., & Kristensson, P. O. (2013, March). Crowdsourcing a hit: measuring workers' pre-task interactions on microtask markets. In *First AAAI Conference on Human Computation and Crowdsourcing*.

Jaffer, U., John, N. W., & Standfield, N. (2013). Surgical Trainee Opinions in the United Kingdom Regarding a Three-Dimensional Virtual Mentoring Environment (MentorSL) in Second Life: Pilot Study. *JMIR Serious Games*, *1* (1), e2.

Jurjanz, L., Donix, M., Amanatidis, E. C., Meyer, S., Poettrich, K., Huebner, T., et al. (2011). Visual Personal Familiarity in Amnestic Mild Cognitive Impairment. *PLoS ONE*, *6* (5), e20030.

Kobayashi, M., Ishihara, T., Itoko, T., Takagi, H., & Asakawa, C. (2013). Age-based task specialization for crowdsourced proofreading. In *Universal Access in Human-Computer Interaction. User and Context Diversity* (pp. 104-112). Springer Berlin Heidelberg.

Koen, J. D., & Yonelinas, A. P. (2014). The Effects of Healthy Aging, Amnestic Mild Cognitive Impairment, and Alzheimer's Disease on Recollection and Familiarity: A Meta-Analytic Review. *Neuropsychological Reviews*, *24*, 332–354.

Leonardi, C., Mennecozzi, C., Not, E., Pianesi, F., & Zancanaro, M. (2008). Designing a Familiar Technology For Elderly People. *Gerontechnology 2008; 7(2):151*, *7* (2), 151.

Li, B., Yu, H., Shen, Z., Cui, L., & Lesser, V. R. (2015). *An Evolutionary Framework for Multi-Agent Organizations.* Paper presented at the 2015 IEEE/WIC/ACM International Joint Conference on Web Intelligence and Intelligent Agent Technology (WI-IAT'15).

Li, B., Yu, H., Shen, Z., & Miao, C. (2009). *Evolutionary organizational search.* Paper presented at the Proceedings of The 8th International Conference on Autonomous Agents and Multiagent Systems-Volume 2.

Li, P., Xiangxu, M., Zhiqi, S., & Han, Y. (2009, December). A reputation pattern for service oriented computing. In *Information, Communications and Signal Processing, 2009. ICICS 2009. 7th International Conference on* (pp. 1-5). IEEE.

Lin, H., Hou, J., Yu, H., Shen, Z., & Miao, C. (2015). *An Agent-based Game Platform for Exercising People's Prospective Memory.* Paper presented at the 2015 IEEE/WIC/ACM International Joint Conference on Web Intelligence and Intelligent Agent Technology (WI-IAT'15).

Lin, J., Miao, C., & Yu, H. (2011). A cloud and agent based architecture design for an educational mobile SNS game *Edutainment Technologies. Educational Games and Virtual Reality/Augmented Reality Applications* (pp. 212-219): Springer Berlin Heidelberg.

Lin, J., Yu, H., Miao, C., & Shen, Z. (2015). *An Affective Agent for Studying Composite Emotions.* Paper presented at the 14th International Conference on Autonomous Agents and Multi-Agent Systems (AAMAS'15).

Lin, J., Yu, H., Shen, Z., & Miao, C. (2014). *Studying Task Allocation Decisions of Novice Agile Teams with Data from Agile Project Management Tools.* Paper presented at the 29th IEEE/ACM International Conference on Automated Software Engineering (ASE'14).

Lin, J., Yu, H., Shen, Z., & Miao, C. (2014). *Using Goal Net to Model User Stories in Agile Software Development.* Paper presented at the 15th IEEE/ACIS International Conference on Software Engineering, Artificial Intelligence, Networking and Parallel/Distributed Computing (SNPD'14).

Liu, S., Miao, C., Liu, Y., Fang, H., Yu, H., Zhang, J., & Leung, C. (2015). *A Reputation Revision Mechanism to Mitigate the Negative Effects of Misreported Ratings.* Paper presented at the 17th International Conference on Electronic Commerce (ICEC'15).

Liu, S., Miao, C., Liu, Y., Yu, H., Zhang, J., & Leung, C. (2015). *An Incentive Mechanism to Elicit Truthful Opinions for Crowdsourced Multiple Choice Consensus Tasks.* Paper presented at the 2015 IEEE/WIC/ACM International Joint Conference on Web Intelligence and Intelligent Agent Technology (WI-IAT'15).

Liu, S., Yu, H., Miao, C., & Kot, A. C. (2013). *A Fuzzy Logic Based Reputation Model Against Unfair Ratings.* Paper presented at the 12th International Conference on Autonomous Agents and Multi-Agent Systems (AAMAS'13).

Liu, Y., Liu, S., Fang, H., Zhang, J., Yu, H., & Miao, C. (2014). *RepRev: Mitigating the Negative Effects of Misreported Ratings.* Paper presented at the 28th AAAI Conference on Artificial Intelligence (AAAI-14).

Liu, Y., Zhang, J., Yu, H., & Miao, C. (2014). *Reputation-aware Continuous Double*

*Auction.* Paper presented at the 28th AAAI Conference on Artificial Intelligence (AAAI-14).

Mei, J.-P., Yu, H., Liu, Y., Shen, Z., & Miao, C. (2014). *A Social Trust Model Considering Trustees' Influence.* Paper presented at the 17th International Conference on Principles and Practice of Multi-Agent Systems (PRIMA'14).

Miao, C., Q Yang, H Fang, and A Goh, "Fuzzy cognitive agents for personalized recommendation Web Information Systems Engineering,", In *Proceedings of the 3rd International Conference on Web Information Systems Engineering (WISE'02)*, 362-371, 2002.

Miao, C., A Goh, Y Miao, and Z Yang, "A dynamic inference model for intelligent agents," International Journal of Software Engineering and Knowledge Engineering 11 (05), 509-528, 2001.

Mitchell, T. J., Chen, S. Y., & Macredie, R. D. (2005). Hypermedia learning and prior knowledge: domain expertise vs. system expertise. *Journal of Computer Assisted Learning , 21* (1), 53-64.

Musha, T., Kimura, S., & Wada, K. (2005). Thusday Morning. *Efficacy of robot therapy evaluated by DIMENSION , 3* (4), 208-210.

Naroditskiy, V., Stein, S., Tonin, M., Tran-Thanh, L., Vlassopoulos, M., & Jennings, N. R. (2014, May). Referral incentives in crowdfunding. In *Second AAAI Conference on Human Computation and Crowdsourcing*.

Pan, L., Meng, X., Shen, Z., & Yu, H. (2009). *A reputation pattern for service oriented computing.* Paper presented at the Information, Communications and Signal Processing, 2009. ICICS 2009. 7th International Conference on.

Pan, L., X Luo, X Meng, C Miao, M He, and X Guo, "A Two-Stage Win–Win Multiattribute Negotiation Model: Optimization and Then Concession," Computational Intelligence 29 (4), 577-626, 2013.

Pan, Z., Miao, C., Tan, B. T. H., Yu, H., & Leung, C. (2015). *Agent Augmented Inter-generational Crowdsourcing.* Paper presented at the 2015 IEEE/WIC/ACM International Joint Conference on Web Intelligence and Intelligent Agent Technology (WI-IAT'15).

Pan, Z., Miao, C., Yu, H., Leung, C., & Chin, J. J. (2015). *The Effects of Familiarity Design on the Adoption of Wellness Games by the Elderly.* Paper presented at the 2015 IEEE/WIC/ACM International Joint Conference on Web Intelligence and Intelligent Agent Technology (WI-IAT'15).

Pan, Z., Yu, H., Miao, C., & Leung, C. (2016). *Efficient Collaborative Crowdsourcing.* Paper presented at the 30th AAAI Conference on Artificial Intelligence (AAAI-16).

Pinto, M. R., Medici, S. D., & Napoli, C. (2000). Ergonomics, gerontechnology and well-being in older patients with cardiovascular disease. *International Journal of Cardiology , 72*, 187–188.

Pinto, M. R., Medici, S. D., Sant, C. V., & Bianchi, A. (2000). Ergonomics, gerontechnology, and design for the home-environment. *Applied Ergonomics , 31*, 317-322.

Qin, T., Yu, H., Leung, C., Shen, Z., & Miao, C. (2009). Towards a trust aware cognitive radio architecture. *ACM Sigmobile Mobile Computing and Communications Review, 13*(2), 86-95.

Shen, Z., Yu, H., Miao, C., Li, S., & Chen, Y. (2016). *Multi-Agent System Development MADE Easy.* Paper presented at the 30th AAAI Conference on Artificial

Intelligence (AAAI-16).

Shen, Z., Yu, H., Miao, C., & Weng, J. (2011). Trust-based web service selection in virtual communities. *Web Intelligence and Agent Systems*, *9*(3), 227-238.

Shi, Y., Sun, C., Li, Q., Cui, L., Yu, H., & Miao, C. (2016). *A Fraud Resilient Medical Insurance Claim System.* Paper presented at the 30th AAAI Conference on Artificial Intelligence (AAAI-16).

Silveira, V. E., Daniel, F., Casati, F., & de Bruin, E. D. (2013). Motivating and assisting physical exercise in independently living older adults: a pilot study. *International Journal of Medical Informatics* , *82* (5), 325-34.

Son, G. R., Therrien, B., & Whall, A. (2002). Implicit Memory and Familiarity Among Elders with Dementia. *Journal of nursing scholarship* , *34* (3), 263-267.

Song, HJ., CY Miao, ZQ Shen, W Roel, DH Maja, C Francky, "Design of fuzzy cognitive maps using neural networks for predicting chaotic time series," Neural Networks 23 (10), 1264-1275, 2010.

Song, HJ., ZQ Shen, CY Miao, Y Miao, and BS Lee, "A fuzzy neural network with fuzzy impact grades," Neurocomputing 72 (13), 3098-3122, 2009.

Sung, M. S., & Chang, M. (2005). Use of preferred music to decrease agitated behaviours in older people with dementia: a review of the literature. *Journal of Clinical Nursing* , *14* (9), 1133–1140.

Tao, X., Shen, Z., Miao, C., Theng, Y.-L., Miao, Y., & Yu, H. (2011). Automated negotiation through a cooperative-competitive model *Innovations in Agent-Based Complex Automated Negotiations* (pp. 161-178): Springer Berlin Heidelberg.

Turner, P. (2008). Being-with: A study of familiarity. *Interacting with Computers* , 447–454.

Turner, P., & van de Walle, G. (2006). Familiarity as a basis for Universal Design. *Gerontechnology* , *5* (3), 150-159.

U. Nations, "World population ageing: 1950-2050," *http://www.un.org/esa/population/publications/worldageing19502050/*, 2002.

Weiermann, B., Stephan, M. A., Kaelin-Lang, A., & Meier, B. (2010). Is there a Recognition Memory Deficit in Parkinson's Disease? Evidence from Estimates of Recollection and Familiarity. *International Journal of Neuroscience* , *120*, 211–216.

Weng, J., C Miao, A Goh, Z Shen, and R Gay, "Trust-based agent community for collaborative recommendation," In Proceedings of the 5th international joint conference on Autonomous agents and multi-agent systems (AAMAS'06), 1260-1262, 2016.

Werner, P., Heinik, J., & Kitai, E. (2013). Familiarity, knowledge, and preferences of family physicians regarding mild cognitive impairment. *International Psychogeriatrics* , *25* (5), 805–813.

Wu, Q., Han, X., Yu, H., Shen, Z., & Miao, C. (2013, May). The innovative application of learning companions in virtual singapura. In *Proceedings of the 2013 international conference on Autonomous agents and multi-agent systems*(pp. 1171-1172). International Foundation for Autonomous Agents and Multiagent Systems.

Yu, H., Cai, Y., Shen, Z., Tao, X., & Miao, C. (2010). *Agents as intelligent user interfaces for the net generation.* Paper presented at the Proceedings of the 15th international conference on Intelligent user interfaces.

Yu, H., Lin, H., Lim, S. F., Lin, J., Shen, Z., & Miao, C. (2015). *Empirical Analysis of Reputation-aware Task Delegation by Humans from a Multi-agent Game.* Paper presented at the 14th International Conference on Autonomous Agents and Multi-Agent Systems (AAMAS'15).

Yu, H., Liu, S., Kot, A. C., Miao, C., & Leung, C. (2011). *Dynamic witness selection for trustworthy distributed cooperative sensing in cognitive radio networks.* Paper presented at the Proceedings of the 13th IEEE International Conference on Communication Technology (ICCT'11).

Yu, H., Miao, C., An, B., Leung, C., & Lesser, V. R. (2013). *A Reputation Management Approach for Resource Constrained Trustee Agents.* Paper presented at the 23rd International Joint Conference on Artificial Intelligence (IJCAI'13).

Yu, H., Miao, C., An, B., Shen, Z., & Leung, C. (2014). *Reputation-aware Task Allocation for Human Trustees.* Paper presented at the 13th International Conference on Autonomous Agents and Multi-Agent Systems (AAMAS'14).

Yu, H., Miao, C., Liu, S., Pan, Z., Khalid, N. S. B., Shen, Z., & Leung, C. (2016). *Productive Aging through Intelligent Personalized Crowdsourcing.* Paper presented at the 30th AAAI Conference on Artificial Intelligence (AAAI-16).

Yu, H., Miao, C., & Shen, Z. (2015). Apparatus and Method for Efficient Task Allocation in Crowdsourcing: US Patent App. 14/656,009.

Yu, H., Miao, C., Shen, Z., & Leung, C. (2015). *Quality and Budget aware Task Allocation for Spatial Crowdsourcing.* Paper presented at the 14th International Conference on Autonomous Agents and Multi-Agent Systems (AAMAS'15).

Yu, H., Miao, C., Shen, Z., Leung, C., Chen, Y., & Yang, Q. (2015). *Efficient Task Sub-delegation for Crowdsourcing.* Paper presented at the 29th AAAI Conference on Artificial Intelligence (AAAI-15).

Yu, H., Miao, C., Tao, X., Shen, Z., Cai, Y., Li, B., & Miao, Y. (2009). *Teachable Agents in Virtual Learning Environments: a Case Study.* Paper presented at the World Conference on E-Learning in Corporate, Government, Healthcare, and Higher Education.

Yu, H., Miao, C., Weng, X., & Leung, C. (2012). *A simple, general and robust trust agent to help elderly select online services.* Paper presented at the Network of Ergonomics Societies Conference (SEANES), 2012 Southeast Asian.

Yu, H., Shen, Z., & An, B. (2012). An Adaptive Witness Selection Method for Reputation-Based Trust Models. *PRIMA 2012: Principles and Practice of Multi-Agent Systems*, 184-198.

Yu, H., Shen, Z., & Leung, C. (2013). Towards Health Care Service Ecosystem Management for the Elderly. *International Journal of Information Technology (IJIT), 19*(2).

Yu, H., Shen, Z., Leung, C., Miao, C., & Lesser, V. R. (2013). A Survey of Multi-agent Trust Management Systems. *IEEE Access, 1*(1), 35-50.

Yu, H., Shen, Z., Li, X., Leung, C., & Miao, C. (2012). Whose Opinions to Trust more, your own or others'? *The 1st Workshop on Incentives and Trust in E-commerce - the 13th ACM Conference on Electronic Commerce (WIT-EC'12)*, 1-12.

Yu, H., Shen, Z., & Miao, C. (2007). *Intelligent software agent design tool using goal net methodology.* Paper presented at the Proceedings of the 2007 IEEE/WIC/ACM International Conference on Intelligent Agent Technology.

Yu, H., Shen, Z., & Miao, C. (2008). A goal-oriented development tool to automate the incorporation of intelligent agents into interactive digital media applications. *Computers in Entertainment (CIE), 6*(2), 24.

Yu, H., Shen, Z., & Miao, C. (2009). *A trustworthy beacon-based location tracking*

*model for body area sensor networks in m-health.* Paper presented at the Information, Communications and Signal Processing, 2009. ICICS 2009. 7th International Conference on.

Yu, H., Shen, Z., Miao, C., & An, B. (2012). *Challenges and Opportunities for Trust Management in Crowdsourcing.* Paper presented at the IEEE/WIC/ACM International Conference on Intelligent Agent Technology (IAT).

Yu, H., Shen, Z., Miao, C., & An, B. (2013). *A Reputation-aware Decision-making Approach for Improving the Efficiency of Crowdsourcing Systems.* Paper presented at the 12th International Conference on Autonomous Agents and Multi-Agent Systems (AAMAS'13).

Yu, H., Shen, Z., Miao, C., An, B., & Leung, C. (2014). Filtering Trust Opinions through Reinforcement Learning. *Decision Support Systems (DSS), 66*, 102-113.

Yu, H., Shen, Z., Miao, C., Leung, C., & Niyato, D. (2010). A survey of trust and reputation management systems in wireless communications. *Proceedings of the IEEE, 98*(10), 1755-1772.

Yu, H., Shen, Z., Miao, C., & Tan, A.-H. (2011). *A simple curious agent to help people be curious.* Paper presented at the 10th International Conference on Autonomous Agents and Multiagent Systems-Volume 3.

Yu, H., Shen, Z., Miao, C., Wen, J., & Yang, Q. (2007). *A service based multi-agent system design tool for modelling integrated manufacturing and service systems.* Paper presented at the Emerging Technologies and Factory Automation, 2007. ETFA. IEEE Conference on.

Yu, H., & Tian, Y. (2005). *Developing Multiplayer Mobile Game Using MIDP 2.0 Game API and JSR-82 Java Bluetooth API.* Paper presented at the 2005 Cybergames Conference.

Yu, H., Yu, X., Lim, S. F., Lin, J., Shen, Z., & Miao, C. (2014). *A Multi-Agent Game for Studying Human Decision-making.* Paper presented at the 13th International Conference on Autonomous Agents and Multi-Agent Systems (AAMAS'14).

Z Man, K Lee, D Wang, Z Cao, and C Miao, "A new robust training algorithm for a class of single-hidden layer feedforward neural networks," Neurocomputing 74 (16), 2491-2501, 2011.

Zhao, G., Z Shen, C Miao, and Z Man, "On improving the conditioning of extreme learning machine: a linear case," In *Proceedings of the 7th International Conference on Information, Communications and Signal Processing (ICICS'09)*, 2009.

Zhao, YZ., M Ma, CY Miao, and TN Nguyen, "An energy-efficient and low-latency MAC protocol with Adaptive Scheduling for multi-hop wireless sensor networks," Computer Communications 33 (12), 1452-1461, 2010.

Zhao, YZ., C Miao, M Ma, JB Zhang, and C Leung, "A survey and projection on medium access control protocols for wireless sensor networks," ACM Computing Surveys (CSUR) 45 (1), 7, 2012.